\documentclass[sigconf,copyrightmode,screen]{acmart}

\usepackage{popets}

\usepackage{threeparttable}
\usepackage{colortbl}
\usepackage{changepage}
\usepackage{setspace}
\newcommand{\eg}{e.g., }

\newcommand{\ie}{i.e., }

\setcopyright{popets}
\copyrightyear{2025}

\acmYear{YYYY}
\acmVolume{YYYY}
\acmNumber{X}
\acmDOI{}
\acmConference{Proceedings on Privacy Enhancing Technologies}
\acmISBN{}
\begin{document}

\title[A Risk Assessment Framework for Digital Identification Systems]{A Risk Assessment Framework for Digital Identification Systems}

\author{Allison Woodruff}
\affiliation{%
  \institution{Google}
  \city{} 
  \state{} 
  \country{} 
}
\email{woodruff@acm.org}

\author{Dirk Balfanz}
\affiliation{%
  \institution{Google}
  \city{} 
  \state{} 
  \country{} 
}
\email{balfanz@google.com}

\author{Will Drewry}
\affiliation{%
  \institution{Google}
  \city{} 
  \state{} 
  \country{} 
}
\email{drewry@google.com}

\author{Mariana Raykova}
\affiliation{%
  \institution{Google}
  \city{} 
  \state{} 
  \country{} 
}
\email{marianar@google.com}

\renewcommand{\shortauthors}{Woodruff et al.}

\begin{abstract}
We introduce a risk assessment framework for digital identification systems, as well as recommended best practices to enhance privacy, security, and other desirable properties in these systems. To generate these resources, we created a casebook of a wide range of digital identification systems, and we then applied expert analysis and critique to identify patterns. We piloted the framework on several reviews within our organization over a period of approximately one year, and found it to be robust and helpful for those reviews. This work is intended to inform product review and development, product policy, and standards efforts, and to help guide a consistent responsible approach to digital identification across the broader digital identification ecosystem.
\end{abstract}

\keywords{digital identification, privacy, risk assessment, security}

\maketitle

\pagestyle{plain}

\section{Introduction}
\label{sec:intro}

In recent years, many digital identification systems have been launched or proposed. For example, the press has reported on refugees scanning their irises to authenticate themselves to receive food and cash in certain refugee camps~\cite{kramer,juskalian,staton}. The company deploying the technology has promoted it as a secure way for refugees to efficiently prove their identities, \eg to quickly check out of a grocery store at a refugee camp~\cite{kramer}. Proponents have also pointed out that iris scans are an effective way to minimize fraud from people who ``dual register'' to get extra benefits~\cite{kramer,staton}. At the same time, refugee advocates have raised substantial privacy and human rights concerns such as creating a database of stable identifiers of vulnerable populations~\cite{kramer,juskalian,staton}. Beyond distribution of goods and services~\cite{economicinclusion,sato}, many other use cases for digital identification systems have emerged as well, \eg mobile driver's licenses~\cite{ca-dmv}, transportation cards~\cite{mta,heydt2006privacy}, health certificates~\cite{beduschi2021,mbunge,technicalcommons}, human verification~\cite{berwick2024,heath2025}, social media verification~\cite{powers}, driver authentication~\cite{driverauth,watkins,watkinsCSCW}, identification of homeless individuals~\cite{vice}, and more.

Such digital identification systems offer many benefits such as convenience and authentication. Further, digital identification offers substantial advantages over physical identity cards, for example, allowing people to present the minimum information necessary rather than all information on their government identity card, \eg digital identification may allow a user to show their age without revealing additional personal information such as their address. At the same time, digital identification may create substantial risks such as impersonation, fraud, surveillance, increased government access to data, or broad data breaches.

Demand for digital versions of government identification, health certificates, and other documentation is growing rapidly~\cite{sato}, accompanied by public policy calls for responsible development of such technology~\cite{european,ftc}. However, many aspects of digital identification are as yet unresolved, and best practices for digital identification have not yet been fully established. Even when best practices have been articulated, they have often not been achieved in practice, for example because of lack of awareness or because improvements rely on cooperation across a large ecosystem of issuers\footnote{We define an \textbf{issuer} as an organization that generates a digital card and provides it to an individual, often through an app a user has installed on their phone or by email. For example, a healthcare provider or pharmacy may issue a vaccine card.}, verifiers\footnote{We define a \textbf{verifier} as any organization that reads the information in the credential. For example, a concert venue may check health certificates as a condition of entry.}, and technology providers.

We introduce a framework of risk analysis and best practices for digital identification, based on expert experience and case analysis. The work is intended to inform product review and development, product policy, and standards efforts, and to help guide a consistent responsible approach to digital identification across the wider digital identification ecosystem.

Digital identification is a broad space. Our exploration centers on issues such as provisioning, managing, and presenting ID/credentials (\eg driver's licenses, health cards) issued by governments/author\-ities or by non-government entities---this may also include, for example, mechanisms to verify or impute government ID (\eg the use of facial recognition to infer a person's legal identity), or the use of non-government identification systems to provide government/social services. While authentication is not our main focus, we have considered it in our analysis and provide select illustrative examples related to authentication. Issues such as the following are beyond the scope of our exploration: labeling identity attributes (\eg assigning gender~\cite{classification}), or personal presentation~\cite{goffman} in digital spaces (\eg social identity management).

Further, in this work, we focus largely on identifying and reducing potential risks to users and society, many of which revolve around privacy and security. Even within this scope, we do not consider this work comprehensive, and beyond that we recognize that other considerations must also be taken into account when assessing digital identification systems. Accordingly, we view the framework to be just one tool among many to consider the safer deployment of technology.

The remainder of the paper is structured as follows. We begin by reviewing relevant background. Next we describe our methodology and introduce the framework. We then provide additional discussion and conclude.
\section{Background}

In this section, we review relevant digital identification systems, products, and standards, and we then turn to risk and harms analysis frameworks. 

\subsection{Digital Identification Systems}

Digital identification is a broad and still developing space.  While digital identities have existed for decades~\cite{whitley2014}, and efforts such as the W3C's work on decentralized identifiers (DIDs)~\cite{W3C} have advanced standardization, much work remains to increase awareness, availability, and consistent application of best practices.

\subsubsection{Legal Digital Identity}

The digitization of legal identity is at various levels of deployment in countries globally. Electronic passports may be considered the first widely deployed digital identity, but their use is primarily for international travel (see~\cite{trulySSI} for a discussion of digital support for ``use cases that normally require a physical passport'', and ~\cite{biometricpassports,bordercontrol} for discussion of considerations and concerns with the use of biometrics in border control and e-passport applications). In the European Union, extensive standardization work has been underway for some time to ratify eIDAS (electronic Identification, Authentication, and Trust Services) 2.0 ~\cite{eu-eidas}. In the US, the deployment of mobile driver's licenses (mDL) has been tested in several jurisdictions~\cite{ca-dmv}. 

Legal digital identity deployments often involve many parties: government agencies, government-approved technology vendors, and consumer technology vendors. For example, in the US, government digital identification is the responsibility of each state's department of motor vehicles (DMV)~\cite{path} which need to meet emerging federal guidance~\cite{nistv4}. These government entities then outsource the technical integration work to approved vendors which then integrate with consumer technology providers through standardized or non-standardized means. Each of these steps provides opportunities for risk assessment, consideration of best practices, and application of standards. However, given the many layers of integration, even identity systems which have been undergoing standardization for years, such as mDL, will have gaps in standards coverage--such as a lack of standards specifying provisioning protocols.

\subsubsection{Consumer Digital Identity}

Consumer digital identity is simply the mapping of a person to their associated data with an online software service.  In the past, usernames and passwords, which acted as a public identifier and a secret knowledge factor, were the ways that consumers managed that association. Many consumer devices now support both secure cryptographic key management and local measurement of biometrics, which enable newer approaches, such as FIDO's Passkeys~\cite{fido-passkey}.

Consumer digital identities are usually purpose-driven, from accessing email services to providing loyalty discounts at a chain of stores. While there are many different standards, there is no obligation to rely on any given standard.  Additionally, the purpose-oriented nature may change the requirements that a service provider holds a consumer to, such as providing a valid credit card or photo of a legal identity document.

\subsubsection{Physical Identity}

Underpinning both legal and consumer identity are the associated person's actual physical identity. Historically, biometrics were collected mainly in support of legal identity, binding paperwork to a physical person. With facial recognition technologies, it has become increasingly common to map from biometrics to legal digital identity. Further, with technological advancements in mobile devices, biometrics now underlay not only legal identity but consumer digital identities as well.

\subsection{Risk and Harms Analysis Frameworks}

Disciplines such as security and privacy have a valuable tradition of sharing guidelines that outline best practices and help identify risks and minimize harm~\cite{cranor,gorski,bonneau,gomez-barrero,matthews2025}. Such frameworks allow experienced practitioners and researchers to synthesize and share their insights, making them more accessible to developers, regulators, advocates, and others. Such frameworks are particularly valuable for emerging technologies and topics, where complex real-world practical challenges may not yet have been widely articulated or discussed.

Frameworks emphasizing risks and harms have become more prominent, particularly with the emergence of responsible innovation and responsible AI approaches~\cite{stahl,weidinger}.
Shelby et al. developed a taxonomy of sociotechnical harms of algorithmic systems, based on a review of research on this topic~\cite{shelby}. 
Warford et al. present a framework that synthesizes research on risk factors for at-risk users~\cite{warford}.
Scheuerman et al. employed a grounded theory approach to develop a framework of harmful content online~\cite{scheuerman}.
Limited frameworks have been provided specifically for risk assessment of digital identification systems, one notable exception being~\cite{nist} that uses six high-level impact categories such as ``personal safety'' or ``civil or criminal violations'' to recommend assurance levels.

Overall, risk and harms frameworks can be used to evaluate and improve technical systems. For example, they may be used during formal processes such as launch reviews or adversarial testing, or more informally during product discussions or design sprints. These frameworks can also drive standards and policy agendas.
\section{Methodology}

To frame our exploration of the space, we constructed a casebook of 37 digital identification products representing a broad range of issues. We constructed the casebook from October 2021 through November 2022, and reviewed and analyzed it from June 2022 through January 2023; we overlapped the casebook construction, review, and analysis so that we could iteratively select informative cases and also incorporate newly emerging issues. We chose the casbeook approach because digital identity issues are highly contextual, so specific cases elucidate important risks and mitigations. To construct the casebook, we consulted internal stakeholders and reviewed external literature to identify a wide variety of real and fictional systems/products (\eg~\cite{amazonone,dellinger,driverauth,irs,joseph,kramer,roblox,totilo,venice,vice,watkins}) and then wrote a pr\'ecis of each.

These cases were reviewed and analyzed by four co-authors who are experts in digital identification systems; amongst these reviewers there is also expertise in security, privacy, AI ethics, health ethics, human-computer interaction, human rights, and related issues. We reviewed and independently scored each case according to (1) risk and (2) whether a technology company should build the product. We also optionally wrote open-ended comments on our analysis of each product.

We then held a series of reconciliation meetings, supplemented by occasional written communication, to understand and/or resolve discrepancies. Most discrepancies were due to different priorities and/or missed risks; in the reconciliation discussions a reviewer often adjusted an outlier score to a higher risk level, which brought scores closer together. Discrepancies were typically not due to different assumptions about the facts of the cases. We took detailed notes throughout the reconciliation, for example, noting best practices that in our experience are often overlooked.

We used an inductive approach to explore emerging themes and common patterns in the cases~\cite{thomas}. Drawing on reflexive thematic analysis approaches~\cite{braun2019, braun2020}, we iteratively developed and refined a list of risk factors that were most salient for each of the digital identification cases. In some instances these risk factors stemmed from our own prior experiences reviewing digital identification products. In other instances, we built on and/or extended existing concepts from security, privacy, and related literatures to develop risk factors, \eg trusted versus non-trusted parties, selective disclosure~\cite{selective,u-prove}, unlinkability~\cite{gomez-barrero, minimization}, revocability~\cite{trulySSI}, on-device processing~\cite{iot,mollah}, shared use~\cite{shared}, notice and consent~\cite{onnotice,contextualintegrity,pathologies,sevenflaws}, and inclusion and equity~\cite{barocas,economicinclusion,worldbank}. We do not claim the list of risk factors is exhaustive; part of our contribution here is to prioritize and select from a larger set of possible considerations.\footnote{For example, usability is a consideration for digital identification systems~\cite{korir} but we do not include an overarching risk factor for it, as in our experience that is often well-covered in other aspects of product development and review. At the same time, we do include risk factors for specific aspects of usability which are highly salient and often problematic in digital identification systems, such as \textbf{shared use} and \textbf{notice and consent}.} Our final list included 22 risk assessment factors, and we then organized those risk factors into five main categories. As we progressed, we also took note of other guidance such as common pitfalls and mitigations for individual risk factors. At the end of the process, we had a casebook with a pr\'ecis and a risk and mitigation analysis for each case, and a broader framework with 22 risk factors, recommended best practices, common pitfalls, mitigations, open questions, and other guidance for digital identification systems.

As an early pilot of the framework, we worked through two live product reviews in detail; gathered feedback from nine experts in security, privacy, at-risk users, law, and public policy; and iterated on the content of the framework. We then applied it to eight additional product reviews with approximately 20 cross-functional stakeholders (representatives from product teams, AI ethics, human rights, legal, privacy, product policy, public policy, and security) within our organization over a period of approximately one year and found it to be robust and helpful for those reviews. The bulk of the reviewing took place throughout 2023 with minor activity continuing into 2024. As noted in Section~\ref{sec:intro}, a wide range of considerations must be taken into account, and our work was used as one tool among many to consider the safer deployment of technology.
\section{Framework}

We begin by presenting our framework, comprised of 22 risk assessment factors, shown in Tables~\ref{table:approach} through ~\ref{table:societal}. 
Each factor is presented with a more desirable ``lower risk'' choice on the lefthand side and a less desirable ``higher risk'' (or unnecessary risk) choice on the righthand side. For example, in Table~\ref{table:approach} a \textbf{trusted parties} approach (lefthand side) is typically lower risk than a \textbf{non-trusted parties} approach (righthand side). In some cases these choices are binary while in others they may represent endpoints in a range of choices. The risk factors are grouped into the following five main categories of risk assessment considerations for digital identification systems which emerged from our analysis: 

\begin{itemize}
\item \textbf{High-Level Approach} (Table~\ref{table:approach})
\item \textbf{Desirable Properties} (Table~\ref{table:properties})
\item \textbf{Recommended Techniques} (Table~\ref{table:techniques})
\item \textbf{Undesirable Technical Consequences} (Table~\ref{table:consequences})
\item \textbf{Societal Impact} (Table~\ref{table:societal})
\end{itemize}

\begin{table*}[th!]

\caption{\textnormal{Risk assessment factors for} \textbf{High-Level Approach} \textnormal{(major design choices). Unless otherwise noted, examples throughout the paper are intended to be abstract and/or fictional rather than representing real-world products.}}

\begin{spacing}{1.1}
\begin{center}

\begin{threeparttable}
\begin{tabular}{ p{23em} p{23em} }

\rowcolor{black}
\multicolumn{2}{|l|}{\textcolor{white}{\textbf{High-Level Approach}}} \\

\cellcolor{gray}{\textcolor{white}{\textbf{Lower Risk}}} & \cellcolor{gray}{\textcolor{white}{\textbf{Higher Risk}}} \\

\toprule

\textbf{Trusted Parties}
\newline
\newline\newline\small\emph{Example: A service to present legal ID will display credentials only to government verifiers such as the TSA}
& 
\textbf{Non-Trusted Parties:} Sharing credentials with non-trusted parties can lead to data proliferation
\newline\newline\small\emph{Example: Many parties (often non-trusted verifiers) request users provide health credentials to access physical spaces such as restaurants, bars, gyms, and more}
\\
\cmidrule{1-2}

\textbf{Effective:} The proposed mechanism is sufficiently accurate and feasible to achieve the desired results
\newline\newline\small\emph{Example: Face Unlock stops some, but not all, authentication attacks}
& 
\textbf{Ineffective:} The proposed mechanism is not sufficiently accurate or feasible to achieve the desired results
\newline\newline\small\emph{Example: Relying on voice identification for authentication in situations where high accuracy is needed}
\\
\cmidrule{1-2}

\textbf{Biometrics Uniquely Advantageous}\tnote{a}~: If biometrics are used, biometrics appear inherently useful and well-suited to the use case; it may be possible to demonstrate that non-biometrics can not achieve the product goal
\newline\newline\small\emph{Example: Use of facial recognition in combination with other credentials for access to high security areas in a workplace}
& 
\textbf{Biometrics Used When Not Necessary:} Biometrics are used but a non-biometric mechanism is available that would be less risky, and equally or more effective
\newline
\newline\newline\small\emph{Example: Use of palm prints at point of purchase; while palm prints appear preferable to face matching, it may further reduce risk to rely on a proxy device or card~\cite{amazonone,palmprint}}
\\
\cmidrule{1-2}

\textbf{Government Actor Performing Government Function}
\newline\newline\small\emph{Example: Government builds an app and verification process to issue and present digital version of legal ID}
& 
\textbf{Non-Government Actor Performing (Pseudo) Government Function}
\newline\newline\small\emph{Example: Third party becomes purveyor of legal ID, or third party verifies user owns a given legal ID}
\\

\bottomrule

\end{tabular}

\begin{tablenotes}
\item[a] \small{Here, when we refer to biometrics, we are considering the case where the product/service provider will verify the biometrics and perhaps store them server side, not just use existing mechanisms like Touch ID. This guidance is not meant to discourage developers from using Touch ID.}
\end{tablenotes}

\end{threeparttable}
\end{center}
\end{spacing}

\label{table:approach}

\end{table*}
\begin{table*}[th!]

\caption{\textnormal{Risk assessment factors for} \textbf{Desirable Properties}\textnormal{.}}

\begin{spacing}{1.1}
\begin{center}
\begin{threeparttable}
\begin{tabular}{ p{23em} p{23em} }

\rowcolor{black}
\multicolumn{2}{|l|}{\textcolor{white}{\textbf{Desirable Properties}}} \\

\cellcolor{gray}{\textcolor{white}{\textbf{Lower Risk}}} & \cellcolor{gray}{\textcolor{white}{\textbf{Higher Risk}}} \\

\toprule

\textbf{Reveals Minimum Information Necessary to Achieve the Goal:} Selective disclosure\tnote{a}
\newline\newline\small\emph{Example: A health app presents vaccination status (\eg up-to-date, not up-to-date) and no further details}\tnote{b}
& 
\textbf{Reveals More Information Than Necessary to Achieve the Goal}
\newline\newline\small\emph{Example: A health app displays all vaccination information such as the person’s name and date/location of vaccinations}
\\
\cmidrule{1-2}

\textbf{Unlinkability:} Cannot link user identity across different uses of the same credential
\newline\newline
\newline\newline\small\emph{Example: An age verification app generates one-time use credentials so that a user’s sessions can not be linked}
& 
\textbf{Links An Identity to Other Data:} Links one identity, such as a legal identity, to other data, such as online activity, OR \textbf{Creates a Record of Transactions Linked to An Identity}
\newline\newline\small\emph{Example: Subway pass and associated transit activity are linked to user account including name, payment information, and other data~\cite{mta}}
\\
\cmidrule{1-2}

\textbf{Revocability:} The issuer can easily revoke and reissue the credential
\newline\newline\small\emph{Example: An app to display a government-issued driver’s license is architected in such a way that the government can revoke the license and/or issue an updated version of the license, \eg with a new expiration date}
& 
\textbf{Not Easily Revocable:} The issuer can not easily revoke and reissue the credential
\newline\newline\small\emph{Example: An app to display a government-issued driver’s license is architected in such a way that the government can not revoke and/or issue an updated version of the license}
\\

\bottomrule

\end{tabular}
\begin{tablenotes}
\item[a] \small{Selective disclosure very often arises in the context of electronic presentation of government-issued credentials, such as government IDs or health certificates.}
\item[b] \small{Consider also the application of selective disclosure for online age verification, where it has been proposed as an improvement over less privacy-conscious online mechanisms~\cite{blazy2024,leffer2024,zero}; at the same time, some argue that in this domain online processes substantially increase risk over in-person presentation even with protections such as selective disclosure in place~\cite{goldman2025,effageverification}.}
\end{tablenotes}
\end{threeparttable}
\end{center}
\end{spacing}

\label{table:properties}

\end{table*}
\begin{table*}[pth!]

\caption{\textnormal{Risk assessment factors for} \textbf{Recommended Techniques}\textnormal{.}} 

\begin{spacing}{1.1}
\begin{center}
\begin{threeparttable}
\begin{tabular}{ p{23em} p{23em} }

\rowcolor{black}
\multicolumn{2}{|l|}{\textcolor{white}{\textbf{Recommended Techniques}}} \\
\cellcolor{gray}{\textcolor{white}{\textbf{Lower Risk}}} & \cellcolor{gray}{\textcolor{white}{\textbf{Higher Risk}}} \\

\toprule

\textbf{Ephemeral:} Risk is lower if data is deleted (almost) immediately, even though data deletion is hard to guarantee
\newline\newline\small\emph{Example: Biometric data is gathered when user is first issued a certificate, and then it is deleted; after issuance non-biometric credentials are used}
& 
\textbf{Permanent Record:} Sharing and holding data carries risk; stable identifiers are more risky than non-stable identifiers
\newline\newline\small\emph{Example: Biometric data is gathered for authentication and then stored permanently}
\\
\cmidrule{1-2}

\textbf{On Device}
\newline\newline\small\emph{Example: The use of facial recognition for unlocking a user’s phone can often be performed on device, greatly reducing potential access to biometric data}
& 
\textbf{Server Side}
\newline\newline\small\emph{Example: Users can use their palm prints at point of purchase, enabled by a server-side database of palm prints~\cite{amazonone,palmprint}}
\\
\cmidrule{1-2}

\textbf{Initial Launch Incorporates All Privacy and Security Best Practices}
\newline\newline\small\emph{Example: Initial launch is delayed to implement a stronger security mechanism}
& 
\textbf{Initial Launch Does Not Incorporate All Privacy and Security Best Practices}
\newline\newline\small\emph{Example: Initial launch includes a reasonable security protection mechanism, but not the strongest standard option available}
\\
\cmidrule{1-2}

\textbf{Effective Protection for Shared Use:} System thoughtfully supports multiple users/accounts per device; supports ID/credentials for minors
\newline
\newline\newline\small\emph{Example: An app to present government ID provides separate profiles for different family members, and allows parents to manage ID for their children}
& 
\textbf{Does Not Have Effective Protection for Shared Use:} System does not sufficiently account for more than one user/account per device; does not support ID/credentials for minors
\newline\newline\small\emph{Example: An app to present government ID allows ID from multiple legal names per account but does not offer security protections for the different identities}
\\
\cmidrule{1-2}

\textbf{Notice/Consent Provided:} Subjects are notified, provided meaningful consent, and have a viable opt-out\tnote{a}
\newline
\newline\newline\small\emph{Example: Meeting attendees are offered the opportunity to opt-in to a system that identifies speakers and creates a transcript with names}
& 
\textbf{Notice/Consent Not Provided:} Subjects are not notified, are not provided meaningful consent, and/or do not have a viable opt-out
\newline\newline\small\emph{Example: People are observed in a public place for urban planning purposes, but are not made aware of ongoing identification and/or are not offered an opt-out}\tnote{b}
\\
\cmidrule{1-2}

\textbf{A Plan is In Place for Outages and Failures:} Users will be well-supported in the event of outages and failures
\newline
\newline\newline\small\emph{Example: Reasonable service guarantees are established}
& 
\textbf{A Plan is Not In Place for Outages and Failures:} Users will need to rely on physical documents or expend undue effort in the event of outages and failures
\newline\newline\small\emph{Example: Users of a digital ID presentation app are told to carry their physical ID with them at all times}\tnote{c}
\\

\bottomrule

\end{tabular}
\begin{tablenotes}
\item[a] \small{We note that as discussed in the literature, while notice and consent can be remediating factors, they may not fully ameliorate risks~\cite{onnotice,contextualintegrity,pathologies}.}
\item[b] \small{While details are not specified in the article, notice/consent issues may pertain in a situation such as the one posed in~\cite{venice}.}
\item[c] \small{Conversely, consider that in some situations it may be advisable to use physical ID rather than digital ID~\cite{delvalle}.}
\end{tablenotes}
\end{threeparttable}
\end{center}
\end{spacing}

\label{table:techniques}

\end{table*}
\begin{table*}[th!]

\caption{\textnormal{Risk assessment factors for} \textbf{Undesirable Technical Consequences}\textnormal{.}} 

\begin{spacing}{1.1}
\begin{center}
\begin{threeparttable}
\begin{tabular}{ p{23em} p{23em} }

\rowcolor{black}
\multicolumn{2}{|l|}{\textcolor{white}{\textbf{Undesirable Technical Consequences}}} \\

\cellcolor{gray}{\textcolor{white}{\textbf{Lower Risk}}} & \cellcolor{gray}{\textcolor{white}{\textbf{Higher Risk}}} \\

\toprule

\textbf{Does Not Significantly Increase Risk of Fraud/Abuse:} System is resistant to fraud/abuse, does not create new threats, and/or reduces pre-existing threats
\newline\newline\small\emph{Example: Rolling out smart cards for credit cards reduced existing fraud (detection of which previously relied on risk models)}
& 
\textbf{Facilitates Fraud/Abuse:} System is easily circumvented and/or creates new threats
\newline
\newline\newline\small\emph{Example: Third party gathers biometric data which if breached could in theory be synthesized or replayed at other sites in order to gain access or privileges}
\\
\cmidrule{1-2}

\textbf{Data Set Already Exists:} Risk is lower if the data set already exists, particularly if it is already public
\newline\newline
\newline\newline\small\emph{ Example: Use of public images of celebrities to identify/prevent celebrity impersonation on social media sites}
& 
\textbf{Constructs a Biometric Data Set:} System creates a new biometric data set; concerning even if held privately by trusted party, and even more concerning if publicly available or held by non-trusted party
\newline\newline\small\emph{Example: Creating a new biometric data set to facilitate provision of services}
\\
\cmidrule{1-2}

\textbf{Verifier Already Has Subject’s Information:} Verifier already knows the information being gathered\tnote{a}
\newline\newline\small\emph{Example: Travel authority already knows that a user with a given legal name has purchased a ticket from one  location to another, and is simply verifying that the person holding the ticket has that legal identity}
& 
\textbf{Verifier Acquires Subject’s Information}
\newline
\newline\newline\small\emph{Example: A bartender checking a person’s government ID to confirm their age does not previously have access to their legal name, address, or other information shown on their ID}
\\

\bottomrule

\end{tabular}
\begin{tablenotes}
\item[a] \small{In many workplace scenarios, or scenarios with high physical security, the verifier already has the subject’s information.}
\end{tablenotes}
\end{threeparttable}
\end{center}
\end{spacing}

\label{table:consequences}

\end{table*}
\begin{table*}[pth!]

\caption{\textnormal{Risk assessment factors for} \textbf{Societal Impact}\textnormal{.}} 

\begin{spacing}{1.1}
\begin{center}
\begin{threeparttable}
\begin{tabular}{ p{23em} p{23em} }

\rowcolor{black}
\multicolumn{2}{|l|}{\textcolor{white}{\textbf{Societal Impact}}} \\

\cellcolor{gray}{\textcolor{white}{\textbf{Lower Risk}}} & \cellcolor{gray}{\textcolor{white}{\textbf{Higher Risk}}} \\

\toprule

\textbf{Misidentification Does Not Cause Substantial Harm\tnote{a}}
\newline
\newline\newline\small\emph{Example: Identify individuals for the purpose of computing aggregate metrics about space usage in the workplace and then discard the identifications}
& 
\textbf{Misidentification May Cause Substantial Harm:} False positives/negatives may result in substantial harm
\newline\newline\small\emph{Example: A facial recognition system that identifies homeless individuals~\cite{vice} might mistakenly label an individual as ``assaultive'' resulting in dangerous confrontation with law enforcement}
\\
\cmidrule{1-2}

\textbf{Surveillance of Specific Individuals for Specific Reasons}
\newline
\newline\newline\small\emph{Example: In a system to monitor financial transactions, identification is made only for transactions above a large mandatory reporting amount}\normalsize\tnote{b}

& 
\textbf{Surveillance of Broad Population:} Broad surveillance without specific evidence against all individuals being surveilled
\newline\newline\small\emph{Example: Automatically connecting to a government resource (\eg each time a user wants to present their ID to a verifier) could leak device information, IP address, or other data from all users}\normalsize\tnote{c}
\\
\cmidrule{1-2}

\textbf{Follows Best Practices for Inclusion and Equity:} Product experience is inclusive and/or expected impacts are equitable
\newline\newline\small\emph{Example: Identification is roughly equally accurate for members of different groups}
& 
\textbf{Is Not Inclusive or Equitable:} Product experience is not inclusive and/or expected impacts are not equitable~\cite{senators,sato}
\newline\newline\small\emph{Example: Has a negative impact on populations without access to high-end devices}
\\
\cmidrule{1-2}

\textbf{Follows Widely Accepted Principles of Human and Civil Rights}\tnote{d}
\newline\newline\small\emph{Example: Vaccine card product undergoes a human rights assessment prior to launch in order to identify and mitigate potential issues}
& 
\textbf{Product Experience/Impacts Do Not Follow Widely Accepted Principles of Human and Civil Rights}
\newline\newline\small\emph{Example: A health app makes it easier to identify certain at-risk groups}\tnote{e}
\\
\cmidrule{1-2}

\textbf{Similar Jurisdictions Are Offered Similar Products and Protections}
\newline\newline\small\emph{Example: A driver’s license product provides the same privacy and security protections to every state in the US}
& 
\textbf{Similar Jurisdictions Are Offered Different Products and Protections} \newline\newline\small\emph{Example: A driver’s license product provides different privacy and security protections to different states in the US}
\\
\cmidrule{1-2}

\textbf{Compelling Benefit:} Use case is compelling, so a full risk/benefit analysis may be warranted
\newline\newline\small\emph{Example: Provide identification for people who are incapacitated}
& 
\textbf{Limited Benefit:} Use case is not compelling, so even minor risk seems unwarranted
\newline\newline\small\emph{Example: Use biometric identification to provide minor customer service experience improvements, such as greeting guests}\tnote{f}
\\

\bottomrule

\end{tabular}
\begin{tablenotes}
\item[a] \small{Here we generally consider misidentifications in which the system makes an inadvertent mistake, although some of these harms could also arise from intentional fraud or abuse.}
\item[b] \small{Consider~\cite{cryptocurrency}.}
\item[c] \small{Consider also use cases discussed in~\cite{gaumond}.}
\item[d] \small{For discussion of human rights considerations in digital identification systems, see~\cite{beduschi2021}.}
\item[e] \small{As another example, consider concerns raised in~\cite{senators}.}
\item[f] \small{Consider~\cite{hotel}.}
\end{tablenotes}
\end{threeparttable}
\end{center}
\end{spacing}

\label{table:societal}

\end{table*}

\subsection{How to Use the Framework}
Here is a typical approach to using the framework: We envision this framework as a list of considerations that reviewers can work through when evaluating a given system, product or proposal. These reviewers may be the developers or designers of a system, perhaps without expertise in digital identification systems. At the same time, given that digital identification is in many ways a fairly nascent space, it is often beneficial to include expert reviewers; these may include cross-functional experts from a range of domains, such as human rights, law, privacy, security,  user experience, and more, working together to share ideas. 

Reviewers can proceed by working through the framework, considering each risk factor in turn.\footnote{We have found it reasonable to work through the risk factors in the order presented, but it is certainly sensible for others to follow whatever order seems most useful for their own purposes.} For each risk factor, reviewers can assess whether the product design follows the ``lower risk'' practice, or if it follows or leans towards the ``higher risk'' practice, or alternatively determine that it is not applicable. If it is applicable and does not follow the ``lower risk'' practice, the reviewers can debate whether it would be possible to adjust the product to follow the more desirable practice, and if so, the pros and cons of making that change.

In some cases it is not viable to follow the lower risk practice. 
For example, certain best practices for privacy and security are more easily followed in some infrastructures than others. Consider on-device processing, which often offers protections that server-side processing does not.\footnote{We do however recognize the limitations of on-device approaches. For example, while on-device approaches protect biometric templates being accessed or stored, they do not protect against physical acquisition of devices.} The architecture of some offerings makes it difficult or infeasible to limit processing to on-device.

Overall, \textbf{if a system does not follow the desirable ``lower risk'' practice, reviewers should robustly consider potential harms and alternative product choices.} This analysis should naturally take into account a wide range of considerations not only from this framework but from many other tools and stakeholders as well, leading to a recommendation whether to move forward as is, or to modify or even cancel the product.

The description above notwithstanding, we hope our readers may use the framework however it best suits their needs, and we imagine it could productively be used as a type of checklist, or modified for use as a scoring system.

\subsection{Example Application of the Framework}

Here we return to an example raised in the introduction, the use of iris scans for authentication for food and cash distribution in refugee camps~\cite{kramer,juskalian,staton,shoemaker}. In practice, one would work through all risk factors in the framework. Here we focus on several of the more salient risk factors to consider for that case:

\begin{itemize}
\item \textbf{Biometrics Used When Not Necessary}. While iris scans offer advantages such as convenience and may open up access to people without devices and/or reduce fraud~\cite{kramer,staton}, lower tech or non-biometric techniques such as a code may be sufficiently effective and incur less risk~\cite{kramer}.
\item \textbf{Constructs a Biometric Dataset}. It is likely the case that a database of iris scans does not previously exist, and constructing a data set of stable, non-alterable identifiers of a sensitive population introduces risk of misuse or security attacks~\cite{kramer,staton}.
\item \textbf{Product Experience and/or Impacts May Not Follow Widely Accepted Principles of Human and Civil Rights}. Given arguments raised by refugee advocates regarding surveillance, consent, and personal safety threats~\cite{kramer,juskalian,staton}, a robust human rights impact assessment should likely be performed to assess possible harms~\cite{staton}.
\end{itemize}

\subsection{Risk Factor Details}

While we do not discuss each risk factor individually for the sake of brevity, many of the risk factors are largely self-explanatory and do not require additional detail. However, some risk factors benefit from additional comments. In this section, we elaborate on more complex risk factors and highlight common pitfalls. Readers may find it useful to begin by reading Tables~\ref{table:approach}\textendash\ref{table:societal}, consulting the details in this section as needed.

\subsubsection{Biometrics Uniquely Advantageous ... Biometrics Used When Not Necessary (Table~\ref{table:approach})}

We recommend assessing whether biometrics offer a genuine advantage for the use case, versus being used unnecessarily. Biometrics are often proposed when less invasive measures can achieve comparable effects. It is perhaps natural for developers to gravitate towards biometric approaches as, for example, they are a convenient way to provide usable security without asking humans to remember a password or other secret. However, biometrics introduce a number of substantial risks, \eg they are often (near) stable identifiers over the lifetime of an individual, so once the information has been gathered, it can enable long-term tracking~\cite{effbiometrics,goodell2019decentralized}; or they may be gathered non-volitionally.\footnote{For example, law enforcement may compel people to unlock their devices with biometrics, even if they can not compel them to provide passwords (\cite{delvalle, fbi}). As another example, face biometrics may be gathered easily in public spaces.} Due to the sensitivity of biometrics, it is often advisable to use alternative mechanisms that provide a level of indirection. It is often wise to scope a system to achieve exactly what is intended and no more, for example, to create a temporary identity that can be used only in a specific situation and not more broadly.

Here is one common pitfall: unnecessarily using biometrics when ``device as proxy'' meets the use case. As the sophistication of consumer devices has improved, it is now possible for those devices to make strong guarantees about access control to cryptographic key material and associated operations. This capability is often leveraged where the device acts as a proxy for the user's identity by registering key material the device holds on behalf of the user and uses only after authenticating them.

Although ``device as proxy'' is already the de-facto standard for many use cases (\eg sign-in, autofill), too often products across the industry fail to adopt this best practice. We recommend guarding against this, and using ``device as proxy'' when practicable.\footnote{In assessing practicality, we recommend considering the bottom tier of devices that can provide such proxy functionality, and assessing whether it meets the use case.} Ever since Touch ID on the iPhone, and perhaps before, it has been a well-known, well-accepted security best practice to gather and keep biometrics on device and then use the device as a proxy to submit something other than biometrics to a server. A typical model is that the user inputs biometrics such as a fingerprint to authenticate themselves on a device in their control, and then the device becomes the interface to other less trusted services. In this way, a user’s device can become a trusted proxy to translate between on-device authentication and external services, bridging between easy-to-use authentication for the user and authentication with desirable properties for online use (\eg cryptographic security and revocability). Some apps use QR codes for this purpose, and for specific use cases (\eg payments, presentation of mobile driver licenses) NFC-based protocols are available.

\subsubsection{Effective ... Ineffective (Table~\ref{table:approach})}

One common pitfall in digital identification systems is to choose an approach that is not feasible for the population/situation, \eg asking rideshare drivers to authenticate themselves while driving, thereby creating hazardous conditions~\cite{watkins}; or incorrectly assuming a given user population has capacity to manage a particular type of secret, account, or device. An advisable mitigation is to engage with community partners and/or conduct user research to confirm that a given approach is appropriate (see Section~\ref{sec:stakeholder} for more discussion of stakeholder engagement).

Another common pitfall is to employ voice and/or affect recognition as a primary factor for identification, a task for which they typically are not sufficiently accurate/secure. Voice has fewer features than facial recognition and is often not performed in a protected environment on device. Further, deep fake technology can make it easy to circumvent voice (see ~\cite{voice} for an example of an AI-generated synthetic voice successfully authenticating a user at a financial institution); voice clips from high profile users may be readily available, or people who have been close to the user can provide data, to relatively easily train a generative model that is sufficiently good to pass voice authentication models. Rather than relying on voice and/or affect recognition as a primary factor, it is advisable to use a different mechanism entirely~\cite{voice}; and/or to use voice, affect recognition, and other lower accuracy identifiers as secondary factors (\ie use them as a component in multi-factor identification).

\subsubsection{Trusted Parties ... Non-Trusted Parties (Table~\ref{table:approach})}

Policies and terms of service can specify how identification systems will be used, and can limit access to trusted parties. For example, access to sensitive technologies such as facial recognition can be controlled via allow lists (see \eg~\cite{celebrity}). While limiting access to trusted parties is ameliorative, and is generally lower risk, one should not assume this is a complete remediation. For example, it is a pitfall to mistakenly assume good actors can keep data secure, as public sector actors (\eg public schools) and others may have not have strong protective measures in place. Relatedly, sometimes an argument is made that it is okay for a trusted actor to execute a potentially problematic practice because the trusted actor will execute it responsibly; however, this is a problematic argument since practices and tools (\eg APIs) that are established and normalized by ``good'' actors can empower ``bad'' actors. We recommend carefully evaluating such assumptions and arguments regarding trusted parties.

\subsubsection{Unlinkability ... Links An Identity to Other Data (Table~\ref{table:properties})}

A system may be designed for unlinkability (see~\eg the approach outlined in~\cite{heydt2006privacy}), such that user identity can not be linked across different uses of the same credential. This affords a number of privacy advantages, and is desirable for many everyday use cases~\cite{goodell2019decentralized}. In fact, Goodell and Aste argue that ``support for multiple, unlinkable identities is an essential right''~\cite{goodell2019decentralized}. Unlinkability can be supported by generating one-time use versions of the credential, and/or by using credential formats and methods that support unlinkability, \eg generating credentials that do not re-use key material, time stamps (credential expiration, etc.), or other correlatable data across credential presentations.

However, unlinkability may be at odds with certain types of personalization or other potential benefits, as it is in opposition to recognizing the same user at a repeat visit. Some systems may therefore support linkability, \eg creating a record of transactions linked to an identity. Using an anonymized identity does not provide full protection in this case, as the record of transactions may reveal deanonymizing or otherwise private information.

\subsubsection{Revocability ... Not Easily Revocable (Table~\ref{table:properties})}

An often overlooked complexity in issuing digital identity and credentials is that they may expire (\eg consider expiration dates on driver’s licenses) or be revoked (\eg consider a driver’s license revoked by a state agency). Presentation of physical identity documents often relies on the notion that a human will examine the document in person and notice if it is out-of-date. However, in many use cases digital identity documents will likely not be examined by a human (for example, they may be checked automatically) and/or the expiration date may not be visible on a digital identity document due to selective disclosure. Accordingly, digital identification systems should consider mechanisms to manage the lifecycle of identity and credentials, including renewal and revocation.

In developing a plan to allow issuers to reissue and revoke credentials, it is important to explore considerations such as the following: whether the approach is resilient to malicious users and/or attackers; and whether the approach will leak data about the user. For example, there may be a strong correlation between criminal activity such as a DUI and revoked credential status, so it is advisable to consider privacy-protective mechanisms for revoking credentials.

\subsubsection{Does Not Significantly Increase Risk of Fraud/Abuse ... Facilitates Fraud/Abuse (Table~\ref{table:consequences})}

A given digital identification system may substantially impact the risk of fraud/abuse. For example, it may be easy to circumvent a given system, or it may be easy to use deep fake technology or otherwise mislead or bypass it. Further, for some high profile individuals, much or all of the necessary information to obtain a digital identity credential may be publicly accessible, so additional protections may be appropriate for such individuals.

As another example, a given identification system may introduce a new exploit or line of attack, creating significant new risk of fraud or abuse beyond the product. For example, it may facilitate impersonation in other services, \eg by constructing a high quality voice model that could be used for voice generation. Alternatively, it may allow a fraudulent purveyor to gather identity metadata from many customers and reuse or sell that data,  or the system may be designed such that it can be appropriated to intentionally inflict individual harm.

Another pitfall is that recovery mechanisms can enable identity theft. Recovery mechanisms are a difficult problem and, currently, they are often (one of) the most insecure part(s) of a system; unfortunately, account recovery often depends on less secure information than the primary authentication mechanism. Therefore, recovery mechanisms are frequently targeted for fraud and abuse. One advisable rule of thumb is that the recovery mechanism should be at least as strong as the actual mechanism.

\subsubsection{Surveillance of Specific Individuals for Specific Reasons ... Surveillance of Broad Population (Table~\ref{table:societal})}

Many commenters have observed that digital identification systems pose significant surveillance risks. One common pitfall is choosing an approach that is unnecessarily expansive, \eg gathers highly sensitive data when less sensitive data would work as well. In this situation it is recommended to choose instead a targeted mechanism for the precise use case.

Linkability is also an important consideration for surveillance, and there are often opportunities to limit it. For example, one can implement static rather than dynamic issuance of government ID for presentation (in the dynamic case, a system could pass each request from the verifier to the issuer, posing privacy challenges). As another example, if long-term surveillance is in place (for example, for urban planning) and biometrics are used, such biometrics can be used exclusively to trade (bootstrap) for a code for regular use to avoid long-term use and surveillance with biometrics.

Another often overlooked consideration is the potential for a service from a ``good'' actor to be used in a different way. Think of dual-use of a given technology, an issue raised in responsible AI (see \eg~\cite{Google}), and assess how quickly and easily a provider might switch from providing a service to conducting surveillance. Further, consider whether the product or feature introduces additional risk of search or data requests. For example, a user stopped by law enforcement might be required to use their phone to transmit their mobile driver’s license data. Consider whether there is any risk the phone might then be subject to additional search.
\section{Discussion}

In this section, we discuss some of the overarching considerations regarding privacy, security, and safety that arise across multiple risk factors in the framework.

\subsection{Standards That Enhance Privacy and Security}

Digital identification requires coordination across a large ecosystem of issuers, verifiers, technology providers, and others. It is difficult for a single player to implement a best practice for digital identity in isolation, although in some cases it is possible for a single technology provider to take an existing standard and layer a best practice on top of it. Standards are a good mechanism for coordination but take a long time (typically years) to put in place, and often lag current technology and best practice. To ensure the ecosystem moves towards best practices as efficiently as possible, standards for digital identity are an important area of investment, especially to support desirable properties called out in the framework such as \textbf{selective disclosure}, \textbf{unlinkability}, and \textbf{revocability}.

As an illustrative example, consider one of the most valuable affordances of digital identification, selective disclosure, \ie the ability to present only the minimal information required in a particular circumstance. For example, when presenting a physical driver’s license to prove their age in order to purchase alcohol at a bar, a person typically shows the front of their driver’s license which includes not only their age but also personal information such as their home address. By contrast, with digital identification, that same person could show only their age along with a photo.\footnote{We note that, while highly protective, selective disclosure may not offer complete privacy protection if credentials with selective disclosure have opaque but persistent, unique value. It is advisable to implement and advocate for standards for single-use credentials with opaque attributes that appear different across different instances of the user’s credential.} However, in practice, selective disclosure is often not used. For example, electronic presentation of proof of COVID-19 vaccination has typically involved display of full information about vaccine administration (type, number, date) from which sensitive information such as age or medical status could be inferred. A more security-conscious implementation would simply confirm whether or not the person is up-to-date on their vaccinations~\cite{technicalcommons}.

Individual actors can not easily establish or enforce selective disclosure, as it must be implemented across issuers, verifiers, and presentation apps; rather this technique is more effectively enforced through cooperative agreements and standards. Standards bodies should more aggressively pursue this important strategy.

\subsection{Launch Velocity}

Companies may experience significant pressure to bring digital identification systems or products to market quickly. For example, they may feel time pressure to bring a partially developed product to market to ``hold their place'' in a rapidly changing landscape, even if current functionality offers only limited user benefit.
Further, they may favor the popular launch-and-iterate innovation strategy that emphasizes agility and continuous improvement based on releasing products with reasonable (but perhaps not ideal) functionality to get real-world experience and feedback. 

These pressures notwithstanding, in this domain it is generally inadvisable to prioritize launch velocity over best practices in privacy, security, fairness, human rights, and other values. The launch-and-iterate innovation strategy founders when actions with serious consequences can not be undone. In high risk or sensitive domains such as digital identity, real-world experimentation can be risky. If something harmful is launched, it may not be possible to roll it back. Key data may already have been exposed, or harmful precedent may have been set for non-ideal practices that other companies then emulate. 

In the digital identification space, it is advisable to move thoughtfully and carefully, minimizing unnecessary exposure by drawing as much as possible on well-established practices such as adversarial testing. We also note that prioritizing launch velocity often creates false dilemmas that can be resolved by thoughtful design and/or additional engineering work. At the same time, ``launching and iterating'' and ``launching safely'' are not entirely at odds. Carefully designed product roadmaps that lay out a path for the next several launches, accompanied by principled protections for each launch, can go a long way towards mitigating potential hazards of an incremental and agile launch process.

\subsection{Stakeholder Engagement}
\label{sec:stakeholder}

Digital identification systems raise many complex sociotechnical issues.
For example, equity is a key consideration for identity, particularly where government-issued credentials are concerned. Accessibility and fairness metrics (for example, measuring performance for different groups) are now increasingly common during product development and launch review. Analyses should also go beyond these issues to consider broadly who is vulnerable and how the product might affect them, for example, children, elders, survivors of intimate partner abuse, refugees, and others. For example, are recovery mechanisms equally accessible to different groups, \eg do in-person recovery mechanisms support urban and rural users equally well? Consider also non-users who may be affected by the product, for example, because it normalizes use of digital identity tools to which they do not have ready access and adds friction to the use of physical documents. Is deployment of the technology likely to create friction for the use of physical documents, disadvantaging those without access to digital identification? 

Such questions are best addressed in partnership with potentially affected communities. Community members, advocates, and regional experts can provide valuable insight regarding real-world impacts of digital identification, including weighing in on speculated harms and proposed mitigations. For example, intimate partner violence scenarios are nuanced and it can be difficult to predict how a given digital identification feature may be (mis)appropriated, especially given complexities with shared devices. Advocates can shed light on harmful scenarios and help prioritize mitigations. 

As a result, stakeholder engagement increasingly features in regulatory perspectives, and robust community partnerships are highly advisable to meet these expectations and thoughtfully address sociotechnical issues.
\section{Conclusions}

We introduced a risk assessment framework for digital identification systems, as well as best practices for them. It is intended as a step towards systematizing knowledge in this emergent and complex domain, drawing on expert experience and assessment~\cite{tenthings} to prioritize among a large set of possible considerations and to highlight certain issues that in our experience are often overlooked. In presenting these resources, we hope to provide utility across a spectrum of users, from developers and regulators to those building agentic systems for process automation to ensure digital identification products are both effective and beneficial.

Beyond the work we have presented here, this space is rife with opportunity for new innovations, standards, and best practices. For example, the technology industry has not yet produced standard protection mechanisms for shared devices or accounts, an issue which interacts heavily with digital identification systems. While this is a challenging problem, it is well worth making an effort to advance the state of the art in this area.


\begin{acks}
We thank Elie Bursztein, Emily Everett, Jen Gennai, Vi Vien Hoang, Patrick Gage Kelley, Kurt Thomas, Amanda Walker, Lawrence You, and our other excellent colleagues at Google for their valuable collaboration, insights, and feedback on this work.
\end{acks}

\bibliographystyle{ACM-Reference-Format}
\bibliography{ourbib}


\begin{thebibliography}{84}


\ifx \showCODEN    \undefined \def \showCODEN     #1{\unskip}     \fi
\ifx \showDOI      \undefined \def \showDOI       #1{#1}\fi
\ifx \showISBNx    \undefined \def \showISBNx     #1{\unskip}     \fi
\ifx \showISBNxiii \undefined \def \showISBNxiii  #1{\unskip}     \fi
\ifx \showISSN     \undefined \def \showISSN      #1{\unskip}     \fi
\ifx \showLCCN     \undefined \def \showLCCN      #1{\unskip}     \fi
\ifx \shownote     \undefined \def \shownote      #1{#1}          \fi
\ifx \showarticletitle \undefined \def \showarticletitle #1{#1}   \fi
\ifx \showURL      \undefined \def \showURL       {\relax}        \fi
\providecommand\bibfield[2]{#2}
\providecommand\bibinfo[2]{#2}
\providecommand\natexlab[1]{#1}
\providecommand\showeprint[2][]{arXiv:#2}

\bibitem[tec(2021)]%
        {technicalcommons}
 \bibinfo{year}{2021}\natexlab{}.
\newblock \bibinfo{title}{{Privacy-Preserving Medical Credentials for Access Authorization}}.
\newblock \bibinfo{howpublished}{Technical Disclosure Commons}.
\newblock
\newblock
\shownote{\url{https://www.tdcommons.org/dpubs_series/4334/}}.


\bibitem[Alliance(nd)]%
        {fido-passkey}
\bibfield{author}{\bibinfo{person}{{FIDO} Alliance}.} \bibinfo{year}{[n.d.]}\natexlab{}.
\newblock \bibinfo{title}{{Passkeys: Accelerating the Availability of Simpler, Stronger Passwordless Sign-Ins}}.
\newblock
\newblock
\newblock
\shownote{\url{https://fidoalliance.org/passkeys/}}.


\bibitem[Alwarafy et~al\mbox{.}(2021)]%
        {iot}
\bibfield{author}{\bibinfo{person}{Abdulmalik Alwarafy}, \bibinfo{person}{Khaled~A. Al-Thelaya}, \bibinfo{person}{Mohamed Abdallah}, \bibinfo{person}{Jens Schneider}, {and} \bibinfo{person}{Mounir Hamdi}.} \bibinfo{year}{2021}\natexlab{}.
\newblock \showarticletitle{{A Survey on Security and Privacy Issues in Edge-Computing-Assisted Internet of Things}}.
\newblock \bibinfo{journal}{\emph{{IEEE Internet of Things Journal}}} \bibinfo{volume}{8}, \bibinfo{number}{6} (\bibinfo{year}{2021}), \bibinfo{pages}{4004--4022}.
\newblock
\urldef\tempurl%
\url{https://doi.org/10.1109/JIOT.2020.3015432}
\showDOI{\tempurl}


\bibitem[Barnes and Schwartz(2019)]%
        {celebrity}
\bibfield{author}{\bibinfo{person}{Parker Barnes} {and} \bibinfo{person}{Andrew Schwartz}.} \bibinfo{year}{2019}\natexlab{}.
\newblock \bibinfo{title}{{Celebrity Recognition now available to approved media \& entertainment customers}}.
\newblock \bibinfo{howpublished}{Google Cloud Blog}.
\newblock
\newblock
\shownote{\url{https://cloud.google.com/blog/products/ai-machine-learning/celebrity-recognition-now-available-to-approved-media-entertainment-customers}}.


\bibitem[Barocas and Nissenbaum(2009)]%
        {onnotice}
\bibfield{author}{\bibinfo{person}{Solon Barocas} {and} \bibinfo{person}{Helen Nissenbaum}.} \bibinfo{year}{2009}\natexlab{}.
\newblock \showarticletitle{{On Notice: The Trouble with Notice and Consent}}. In \bibinfo{booktitle}{\emph{Proceedings of the Engaging Data Forum: The First International Forum on the Application and Management of Personal Electronic Information}}.
\newblock
\newblock
\shownote{\url{https://ssrn.com/abstract=2567409}}.


\bibitem[Barocas and Selbst(2016)]%
        {barocas}
\bibfield{author}{\bibinfo{person}{Solon Barocas} {and} \bibinfo{person}{Andrew~D. Selbst}.} \bibinfo{year}{2016}\natexlab{}.
\newblock \showarticletitle{{Big Data's Disparate Impact}}.
\newblock \bibinfo{journal}{\emph{California Law Review}} \bibinfo{volume}{104}, \bibinfo{number}{3} (\bibinfo{date}{June} \bibinfo{year}{2016}), \bibinfo{pages}{671--732}.
\newblock
\newblock
\shownote{\url{https://www.jstor.org/stable/24758720}}.


\bibitem[Beduschi(2021)]%
        {beduschi2021}
\bibfield{author}{\bibinfo{person}{Ana Beduschi}.} \bibinfo{year}{2021}\natexlab{}.
\newblock \showarticletitle{Rethinking digital identity for post-COVID-19 societies: Data privacy and human rights considerations}.
\newblock \bibinfo{journal}{\emph{Data \& Policy}}  \bibinfo{volume}{3} (\bibinfo{year}{2021}), \bibinfo{pages}{e15}.
\newblock
\urldef\tempurl%
\url{https://doi.org/10.1017/dap.2021.15}
\showDOI{\tempurl}


\bibitem[Berwick(2024)]%
        {berwick2024}
\bibfield{author}{\bibinfo{person}{Angus Berwick}.} \bibinfo{year}{2024}\natexlab{}.
\newblock \showarticletitle{{Sam Altman’s Worldcoin Is Battling With Governments Over Your Eyes}}.
\newblock \bibinfo{journal}{\emph{The Wall Street Journal}} (\bibinfo{date}{18 Aug.} \bibinfo{year}{2024}).
\newblock
\newblock
\shownote{\url{https://www.wsj.com/tech/sam-altman-openai-humanness-iris-scanning-4d0e1dab}}.


\bibitem[Blazy(2024)]%
        {blazy2024}
\bibfield{author}{\bibinfo{person}{Olivier Blazy}.} \bibinfo{year}{2024}\natexlab{}.
\newblock \bibinfo{title}{{Online Age Verification and Privacy Protection: An Impossible Equation?}}
\newblock \bibinfo{howpublished}{Stanford Cyber Policy Center Seminar}.
\newblock
\newblock
\shownote{\url{https://cyber.fsi.stanford.edu/events/may-7-online-age-verification-and-privacy-protection-impossible-equation}}.


\bibitem[Bonneau et~al\mbox{.}(2012)]%
        {bonneau}
\bibfield{author}{\bibinfo{person}{Joseph Bonneau}, \bibinfo{person}{Cormac Herley}, \bibinfo{person}{Paul C.~van Oorschot}, {and} \bibinfo{person}{Frank Stajano}.} \bibinfo{year}{2012}\natexlab{}.
\newblock \showarticletitle{The Quest to Replace Passwords: A Framework for Comparative Evaluation of Web Authentication Schemes}. In \bibinfo{booktitle}{\emph{2012 IEEE Symposium on Security and Privacy}}. \bibinfo{pages}{553--567}.
\newblock
\urldef\tempurl%
\url{https://doi.org/10.1109/SP.2012.44}
\showDOI{\tempurl}


\bibitem[Braun and Clarke(2019)]%
        {braun2019}
\bibfield{author}{\bibinfo{person}{Virginia Braun} {and} \bibinfo{person}{Victoria Clarke}.} \bibinfo{year}{2019}\natexlab{}.
\newblock \showarticletitle{Reflecting on reflexive thematic analysis}.
\newblock \bibinfo{journal}{\emph{Qualitative Research in Sport, Exercise and Health}} \bibinfo{volume}{11}, \bibinfo{number}{4} (\bibinfo{year}{2019}), \bibinfo{pages}{589--597}.
\newblock
\urldef\tempurl%
\url{https://doi.org/10.1080/2159676X.2019.1628806}
\showDOI{\tempurl}


\bibitem[Braun and Clarke(2021)]%
        {braun2020}
\bibfield{author}{\bibinfo{person}{Virginia Braun} {and} \bibinfo{person}{Victoria Clarke}.} \bibinfo{year}{2021}\natexlab{}.
\newblock \showarticletitle{One size fits all? {W}hat counts as quality practice in (reflexive) thematic analysis?}
\newblock \bibinfo{journal}{\emph{Qualitative Research in Psychology}} \bibinfo{volume}{18}, \bibinfo{number}{3} (\bibinfo{year}{2021}), \bibinfo{pages}{328--352}.
\newblock
\urldef\tempurl%
\url{https://doi.org/10.1080/14780887.2020.1769238}
\showDOI{\tempurl}


\bibitem[Brewster(2021)]%
        {fbi}
\bibfield{author}{\bibinfo{person}{Thomas Brewster}.} \bibinfo{year}{2021}\natexlab{}.
\newblock \showarticletitle{{FBI: You Don’t Have To Tell Us Which Body Part Unlocks Your Smartphone—But We Will Guess Anyway}}.
\newblock \bibinfo{journal}{\emph{Forbes}} (\bibinfo{date}{Nov.} \bibinfo{year}{2021}).
\newblock
\newblock
\shownote{\url{https://www.forbes.com/sites/thomasbrewster/2021/11/29/fbi-no-consent-required-for-forced-phone-unlocks-via-finger-or-face/}}.


\bibitem[Bubola(2021)]%
        {venice}
\bibfield{author}{\bibinfo{person}{Emma Bubola}.} \bibinfo{year}{2021}\natexlab{}.
\newblock \showarticletitle{Venice, Overwhelmed by Tourists, Tries Tracking Them}.
\newblock \bibinfo{journal}{\emph{The New York Times}} (\bibinfo{date}{Oct.} \bibinfo{year}{2021}).
\newblock
\newblock
\shownote{\url{https://www.nytimes.com/2021/10/04/world/europe/venice-tourism-surveillance.html}}.


\bibitem[Chen(2021)]%
        {roblox}
\bibfield{author}{\bibinfo{person}{Chris~Aston Chen}.} \bibinfo{year}{2021}\natexlab{}.
\newblock \bibinfo{title}{{Introducing Age Verification}}.
\newblock
\newblock
\newblock
\shownote{\url{https://blog.roblox.com/2021/09/introducing-age-verification/}}.


\bibitem[Commission(nda)]%
        {eu-eidas}
\bibfield{author}{\bibinfo{person}{European Commission}.} \bibinfo{year}{[n.d.]}\natexlab{a}.
\newblock \bibinfo{title}{{Discover eIDAS}}.
\newblock
\newblock
\newblock
\shownote{\url{https://digital-strategy.ec.europa.eu/en/policies/discover-eidas}}.


\bibitem[Commission(ndb)]%
        {european}
\bibfield{author}{\bibinfo{person}{European Commission}.} \bibinfo{year}{[n.d.]}\natexlab{b}.
\newblock \bibinfo{title}{{European Digital Identity}}.
\newblock
\newblock
\newblock
\shownote{\url{https://commission.europa.eu/strategy-and-policy/priorities-2019-2024/europe-fit-digital-age/european-digital-identity_en}}.


\bibitem[Commission(2018)]%
        {ftc}
\bibfield{author}{\bibinfo{person}{Federal~Trade Commission}.} \bibinfo{year}{2018}\natexlab{}.
\newblock \bibinfo{title}{{FTC} Warns About Misuses of Biometric Information and Harm to Consumers}.
\newblock
\newblock
\newblock
\shownote{\url{https://www.ftc.gov/news-events/news/press-releases/2023/05/ftc-warns-about-misuses-biometric-information-harm-consumers}}.


\bibitem[Cox(2022)]%
        {vice}
\bibfield{author}{\bibinfo{person}{Joseph Cox}.} \bibinfo{year}{2022}\natexlab{}.
\newblock \showarticletitle{Tech Firm Offers Cops Facial Recognition to {ID} Homeless People}.
\newblock \bibinfo{journal}{\emph{Vice}} (\bibinfo{date}{Feb.} \bibinfo{year}{2022}).
\newblock
\newblock
\shownote{\url{https://www.vice.com/en/article/wxdp7x/tech-firm-facial-recognition-homeless-people-odin}}.


\bibitem[Cox(2023)]%
        {voice}
\bibfield{author}{\bibinfo{person}{Joseph Cox}.} \bibinfo{year}{2023}\natexlab{}.
\newblock \showarticletitle{How I Broke Into a Bank Account With an AI-Generated Voice}.
\newblock \bibinfo{journal}{\emph{Vice}} (\bibinfo{date}{Feb.} \bibinfo{year}{2023}).
\newblock
\newblock
\shownote{\url{https://www.vice.com/en/article/how-i-broke-into-a-bank-account-with-an-ai-generated-voice/}}.


\bibitem[Cranor(2008)]%
        {cranor}
\bibfield{author}{\bibinfo{person}{Lorrie~Faith Cranor}.} \bibinfo{year}{2008}\natexlab{}.
\newblock \showarticletitle{A framework for reasoning about the human in the loop}. In \bibinfo{booktitle}{\emph{Proceedings of the 1st Conference on Usability, Psychology, and Security}} (San Francisco, California) \emph{(\bibinfo{series}{UPSEC '08})}. \bibinfo{publisher}{USENIX Association}, Article \bibinfo{articleno}{1}, \bibinfo{numpages}{15}~pages.
\newblock
\newblock
\shownote{\url{https://www.usenix.org/legacy/events/upsec08/tech/full_papers/cranor/cranor.pdf}}.


\bibitem[Dellinger(2019)]%
        {dellinger}
\bibfield{author}{\bibinfo{person}{A.J. Dellinger}.} \bibinfo{year}{2019}\natexlab{}.
\newblock \showarticletitle{{US} prisons are reportedly creating `voice print' databases}.
\newblock \bibinfo{journal}{\emph{engadget}} (\bibinfo{date}{31 Jan.} \bibinfo{year}{2019}).
\newblock
\newblock
\shownote{\url{https://www.engadget.com/2019-01-30-us-prison-voice-print-database.html}}.


\bibitem[Dhamija and Dusseault(2008)]%
        {sevenflaws}
\bibfield{author}{\bibinfo{person}{Rachna Dhamija} {and} \bibinfo{person}{Lisa Dusseault}.} \bibinfo{year}{2008}\natexlab{}.
\newblock \showarticletitle{The Seven Flaws of Identity Management: Usability and Security Challenges}.
\newblock \bibinfo{journal}{\emph{IEEE Security \& Privacy}} \bibinfo{volume}{6}, \bibinfo{number}{2} (\bibinfo{year}{2008}), \bibinfo{pages}{24--29}.
\newblock
\urldef\tempurl%
\url{https://doi.org/10.1109/MSP.2008.49}
\showDOI{\tempurl}


\bibitem[(EFF)(nda)]%
        {effageverification}
\bibfield{author}{\bibinfo{person}{Electronic Frontier~Foundation (EFF)}.} \bibinfo{year}{[n.d.]}\natexlab{a}.
\newblock \bibinfo{title}{Age Verification Harms Users of All Ages}.
\newblock
\newblock
\newblock
\shownote{\url{https://www.eff.org/document/age-verification-harms-users-all-ages}}.


\bibitem[(EFF)(ndb)]%
        {effbiometrics}
\bibfield{author}{\bibinfo{person}{Electronic Frontier~Foundation (EFF)}.} \bibinfo{year}{[n.d.]}\natexlab{b}.
\newblock \bibinfo{title}{Biometrics}.
\newblock
\newblock
\newblock
\shownote{\url{https://www.eff.org/issues/biometrics}}.


\bibitem[Gaumond(2021)]%
        {gaumond}
\bibfield{author}{\bibinfo{person}{Eve Gaumond}.} \bibinfo{year}{2021}\natexlab{}.
\newblock \bibinfo{title}{{Artificial Intelligence Act: What Is the European Approach for AI?}}
\newblock \bibinfo{howpublished}{Lawfare}.
\newblock
\newblock
\shownote{\url{https://www.lawfaremedia.org/article/artificial-intelligence-act-what-european-approach-ai}}.


\bibitem[Goffman(1959)]%
        {goffman}
\bibfield{author}{\bibinfo{person}{Erving Goffman}.} \bibinfo{year}{1959}\natexlab{}.
\newblock \bibinfo{booktitle}{\emph{{The Presentation of Self in Everyday Life}}}.
\newblock \bibinfo{publisher}{Bantam Doubleday Dell Publishing Group}.
\newblock


\bibitem[Goldman(2025)]%
        {goldman2025}
\bibfield{author}{\bibinfo{person}{Eric Goldman}.} \bibinfo{year}{2025}\natexlab{}.
\newblock \showarticletitle{{The ``Segregate-and-Suppress''' Approach to Regulating Child Safety Online}}.
\newblock \bibinfo{journal}{\emph{Stanford Technology Law Review}}  \bibinfo{volume}{forthcoming} (\bibinfo{year}{2025}).
\newblock
\newblock
\shownote{\url{https://papers.ssrn.com/sol3/papers.cfm?abstract_id=5208739}}.


\bibitem[Gomez-Barrero et~al\mbox{.}(2018)]%
        {gomez-barrero}
\bibfield{author}{\bibinfo{person}{Marta Gomez-Barrero}, \bibinfo{person}{Javier Galbally}, \bibinfo{person}{Christian Rathgeb}, {and} \bibinfo{person}{Christoph Busch}.} \bibinfo{year}{2018}\natexlab{}.
\newblock \showarticletitle{General Framework to Evaluate Unlinkability in Biometric Template Protection Systems}.
\newblock \bibinfo{journal}{\emph{{IEEE Transactions on Information Forensics and Security}}} \bibinfo{volume}{13}, \bibinfo{number}{6} (\bibinfo{date}{June} \bibinfo{year}{2018}), \bibinfo{pages}{1406--1420}.
\newblock
\urldef\tempurl%
\url{https://doi.org/10.1109/TIFS.2017.2788000}
\showDOI{\tempurl}


\bibitem[Goodell and Aste(2019)]%
        {goodell2019decentralized}
\bibfield{author}{\bibinfo{person}{Geoff Goodell} {and} \bibinfo{person}{Tomaso Aste}.} \bibinfo{year}{2019}\natexlab{}.
\newblock \showarticletitle{{A Decentralized Digital Identity Architecture}}.
\newblock \bibinfo{journal}{\emph{Frontiers in Blockchain}}  \bibinfo{volume}{2} (\bibinfo{year}{2019}), \bibinfo{pages}{17}.
\newblock
\urldef\tempurl%
\url{https://doi.org/10.3389/fbloc.2019.00017}
\showDOI{\tempurl}


\bibitem[Gorski et~al\mbox{.}(2023)]%
        {gorski}
\bibfield{author}{\bibinfo{person}{Peter~Leo Gorski}, \bibinfo{person}{Luigi~Lo Iacono}, {and} \bibinfo{person}{Matthew Smith}.} \bibinfo{year}{2023}\natexlab{}.
\newblock \showarticletitle{{Eight Lightweight Usable Security Principles for Developers}}.
\newblock \bibinfo{journal}{\emph{IEEE Security \& Privacy}} \bibinfo{volume}{21}, \bibinfo{number}{1} (\bibinfo{date}{Jan.} \bibinfo{year}{2023}), \bibinfo{pages}{20--26}.
\newblock
\showISSN{1558-4046}
\urldef\tempurl%
\url{https://doi.org/10.1109/MSEC.2022.3205484}
\showDOI{\tempurl}


\bibitem[Grassi et~al\mbox{.}(2017)]%
        {nist}
\bibfield{author}{\bibinfo{person}{Paul~A. Grassi}, \bibinfo{person}{Michael~E. Garcia}, {and} \bibinfo{person}{James~L. Fenton}.} \bibinfo{year}{2017}\natexlab{}.
\newblock \showarticletitle{{Digital Identity Guidelines}}.
\newblock \bibinfo{journal}{\emph{NIST Special Publication}}  \bibinfo{volume}{800-63-3} (\bibinfo{date}{June} \bibinfo{year}{2017}).
\newblock
\urldef\tempurl%
\url{https://doi.org/10.6028/NIST.SP.800-63-3}
\showDOI{\tempurl}


\bibitem[Group(2016)]%
        {worldbank}
\bibfield{author}{\bibinfo{person}{World~Bank Group}.} \bibinfo{year}{2016}\natexlab{}.
\newblock \bibinfo{title}{{Identification for Development: Strategic Framework}}.
\newblock
\newblock
\newblock
\shownote{\url{https://thedocs.worldbank.org/en/doc/21571460567481655-0190022016/render/April2016ID4DStrategicRoadmapID4D.pdf}}.


\bibitem[Gupta et~al\mbox{.}(2019)]%
        {driverauth}
\bibfield{author}{\bibinfo{person}{Sandeep Gupta}, \bibinfo{person}{Attaullah Buriro}, {and} \bibinfo{person}{Bruno Crispo}.} \bibinfo{year}{2019}\natexlab{}.
\newblock \showarticletitle{{DriverAuth: A risk-based multi-modal biometric-based driver authentication scheme for ride-sharing platforms}}.
\newblock \bibinfo{journal}{\emph{Computers \& Security}}  \bibinfo{volume}{83} (\bibinfo{date}{June} \bibinfo{year}{2019}), \bibinfo{pages}{122--139}.
\newblock
\showISSN{0167-4048}
\urldef\tempurl%
\url{https://doi.org/10.1016/j.cose.2019.01.007}
\showDOI{\tempurl}


\bibitem[Heath(2025)]%
        {heath2025}
\bibfield{author}{\bibinfo{person}{Alex Heath}.} \bibinfo{year}{2025}\natexlab{}.
\newblock \showarticletitle{Sam {A}ltman’s eye-scanning project launches cryptocurrency in the {US}}.
\newblock \bibinfo{journal}{\emph{The Verge}} (\bibinfo{date}{30 April} \bibinfo{year}{2025}).
\newblock
\newblock
\shownote{\url{https://www.theverge.com/cryptocurrency/659011/worldcoin-us-launch-orb-crypto-sam-altman}}.


\bibitem[Heydt-Benjamin et~al\mbox{.}(2006)]%
        {heydt2006privacy}
\bibfield{author}{\bibinfo{person}{Thomas~S. Heydt-Benjamin}, \bibinfo{person}{Hee-Jin Chae}, \bibinfo{person}{Benessa Defend}, {and} \bibinfo{person}{Kevin Fu}.} \bibinfo{year}{2006}\natexlab{}.
\newblock \showarticletitle{Privacy for {P}ublic {T}ransportation}. In \bibinfo{booktitle}{\emph{International Workshop on Privacy Enhancing Technologies}} \emph{(\bibinfo{series}{PET 2006})}. Springer, \bibinfo{pages}{1--19}.
\newblock
\urldef\tempurl%
\url{https://doi.org/10.1007/11957454_1}
\showDOI{\tempurl}


\bibitem[Johnson(2024)]%
        {path}
\bibfield{author}{\bibinfo{person}{Ash Johnson}.} \bibinfo{year}{2024}\natexlab{}.
\newblock \bibinfo{title}{{The Path to Digital Identity in the United States}}.
\newblock
\newblock
\newblock
\shownote{\url{https://itif.org/publications/2024/09/23/path-to-digital-identity-in-the-united-states}}.


\bibitem[Joseph and Nathan(2019)]%
        {joseph}
\bibfield{author}{\bibinfo{person}{George Joseph} {and} \bibinfo{person}{Debbie Nathan}.} \bibinfo{year}{2019}\natexlab{}.
\newblock \showarticletitle{Prisons across the {US} are quietly building databases of incarcerated people’s voice prints}.
\newblock \bibinfo{journal}{\emph{The Intercept}} (\bibinfo{date}{30 Jan.} \bibinfo{year}{2019}).
\newblock
\newblock
\shownote{\url{https://theintercept.com/2019/01/30/prison-voice-prints-databases-securus/}}.


\bibitem[Juskalian(2018)]%
        {juskalian}
\bibfield{author}{\bibinfo{person}{Russ Juskalian}.} \bibinfo{year}{2018}\natexlab{}.
\newblock \showarticletitle{Inside the {J}ordan refugee camp that runs on blockchain}.
\newblock \bibinfo{journal}{\emph{MIT Technology Review}} (\bibinfo{date}{12 April} \bibinfo{year}{2018}).
\newblock
\newblock
\shownote{\url{https://www.technologyreview.com/2018/04/12/143410/inside-the-jordan-refugee-camp-that-runs-on-blockchain/}}.


\bibitem[Korir et~al\mbox{.}(2022)]%
        {korir}
\bibfield{author}{\bibinfo{person}{Maina Korir}, \bibinfo{person}{Simon Parkin}, {and} \bibinfo{person}{Paul Dunphy}.} \bibinfo{year}{2022}\natexlab{}.
\newblock \showarticletitle{{An Empirical Study of a Decentralized Identity Wallet: Usability, Security, and Perspectives on User Control}}. In \bibinfo{booktitle}{\emph{Eighteenth Symposium on Usable Privacy and Security}} \emph{(\bibinfo{series}{SOUPS 2022})}. \bibinfo{pages}{195--211}.
\newblock
\newblock
\shownote{\url{https://usenix.org/conference/soups2022/presentation/korir}}.


\bibitem[Kramer(2021)]%
        {kramer}
\bibfield{author}{\bibinfo{person}{Anna Kramer}.} \bibinfo{year}{2021}\natexlab{}.
\newblock \showarticletitle{Refugees are buying groceries with iris scans. {W}hat could go wrong?}
\newblock \bibinfo{journal}{\emph{protocol}} (\bibinfo{date}{Sept.} \bibinfo{year}{2021}).
\newblock


\bibitem[Kumar(2020)]%
        {amazonone}
\bibfield{author}{\bibinfo{person}{Dilip Kumar}.} \bibinfo{year}{2020}\natexlab{}.
\newblock \showarticletitle{Introducing {A}mazon {O}ne---a new innovation to make everyday activities effortless}.
\newblock \bibinfo{journal}{\emph{Innovation at Amazon}} (\bibinfo{date}{Sept.} \bibinfo{year}{2020}).
\newblock
\newblock
\shownote{\url{https://www.aboutamazon.com/news/innovation-at-amazon/introducing-amazon-one-a-new-innovation-to-make-everyday-activities-effortless}}.


\bibitem[Labati et~al\mbox{.}(2016)]%
        {bordercontrol}
\bibfield{author}{\bibinfo{person}{Ruggero~Donida Labati}, \bibinfo{person}{Angelo Genovese}, \bibinfo{person}{Enrique Mu\~{n}oz}, \bibinfo{person}{Vincenzo Piuri}, \bibinfo{person}{Fabio Scotti}, {and} \bibinfo{person}{Gianluca Sforza}.} \bibinfo{year}{2016}\natexlab{}.
\newblock \showarticletitle{{Biometric Recognition in Automated Border Control: A Survey}}.
\newblock \bibinfo{journal}{\emph{Comput. Surveys}} \bibinfo{volume}{49}, \bibinfo{number}{2}, Article \bibinfo{articleno}{24} (\bibinfo{date}{June} \bibinfo{year}{2016}), \bibinfo{numpages}{39}~pages.
\newblock
\showISSN{0360-0300}
\urldef\tempurl%
\url{https://doi.org/10.1145/2933241}
\showDOI{\tempurl}


\bibitem[Leffer(2024)]%
        {leffer2024}
\bibfield{author}{\bibinfo{person}{Lauren Leffer}.} \bibinfo{year}{2024}\natexlab{}.
\newblock \showarticletitle{{Online Age Verification Laws Could Do More Harm Than Good}}.
\newblock \bibinfo{journal}{\emph{Scientific American}} (\bibinfo{date}{April} \bibinfo{year}{2024}).
\newblock
\newblock
\shownote{\url{https://www.scientificamerican.com/article/online-age-verification-laws-privacy/}}.


\bibitem[Matthews et~al\mbox{.}(2025)]%
        {matthews2025}
\bibfield{author}{\bibinfo{person}{Tara Matthews}, \bibinfo{person}{Elie Bursztein}, \bibinfo{person}{Patrick~Gage Kelley}, \bibinfo{person}{Lea Kissner}, \bibinfo{person}{Andreas Kramm}, \bibinfo{person}{Andrew Oplinger}, \bibinfo{person}{Andreas Schou}, \bibinfo{person}{Manya Sleeper}, \bibinfo{person}{Stephan Somogyi}, \bibinfo{person}{Dalila Szostak}, \bibinfo{person}{Kurt Thomas}, \bibinfo{person}{Anna Turner}, \bibinfo{person}{Jill~Palzkill Woelfer}, \bibinfo{person}{Lawrence~L. You}, \bibinfo{person}{Izzie Zahorian}, {and} \bibinfo{person}{Sunny Consolvo}.} \bibinfo{year}{2025}\natexlab{}.
\newblock \showarticletitle{Supporting the Digital Safety of At-Risk Users: Lessons Learned from 9+ Years of Research \& Training}.
\newblock \bibinfo{journal}{\emph{ACM Transactions on Computer-Human Interaction}} (\bibinfo{date}{Feb.} \bibinfo{year}{2025}).
\newblock
\urldef\tempurl%
\url{https://doi.org/10.1145/3716382}
\showDOI{\tempurl}


\bibitem[Matthews et~al\mbox{.}(2016)]%
        {shared}
\bibfield{author}{\bibinfo{person}{Tara Matthews}, \bibinfo{person}{Kerwell Liao}, \bibinfo{person}{Anna Turner}, \bibinfo{person}{Marianne Berkovich}, \bibinfo{person}{Robert Reeder}, {and} \bibinfo{person}{Sunny Consolvo}.} \bibinfo{year}{2016}\natexlab{}.
\newblock \showarticletitle{{``She'll just grab any device that's closer'': A Study of Everyday Device \& Account Sharing in Households}}. In \bibinfo{booktitle}{\emph{Proceedings of the 2016 CHI Conference on Human Factors in Computing Systems}} (San Jose, California, USA) \emph{(\bibinfo{series}{CHI '16})}. \bibinfo{publisher}{Association for Computing Machinery}, \bibinfo{address}{New York, NY, USA}, \bibinfo{pages}{5921–5932}.
\newblock
\urldef\tempurl%
\url{https://doi.org/10.1145/2858036.2858051}
\showDOI{\tempurl}


\bibitem[Mbunge et~al\mbox{.}(2021)]%
        {mbunge}
\bibfield{author}{\bibinfo{person}{Elliot Mbunge}, \bibinfo{person}{Tafadzwa Dzinamarira}, \bibinfo{person}{Stephen~G. Fashoto}, {and} \bibinfo{person}{John Batani}.} \bibinfo{year}{2021}\natexlab{}.
\newblock \showarticletitle{Emerging technologies and {COVID-19} digital vaccination certificates and passports}.
\newblock \bibinfo{journal}{\emph{Public Health in Practice}} (\bibinfo{date}{May} \bibinfo{year}{2021}).
\newblock
\urldef\tempurl%
\url{https://doi.org/10.1016/j.puhip.2021.100136}
\showDOI{\tempurl}


\bibitem[McNeill(2024)]%
        {senators}
\bibfield{author}{\bibinfo{person}{Zane McNeill}.} \bibinfo{year}{2024}\natexlab{}.
\newblock \showarticletitle{{Democratic Senators Ask DOJ If Facial Recognition Tech Violates Civil Rights Act}}.
\newblock \bibinfo{journal}{\emph{Truthout}} (\bibinfo{date}{23 Jan.} \bibinfo{year}{2024}).
\newblock
\newblock
\shownote{\url{https://truthout.org/articles/democratic-senators-ask-doj-if-facial-recognition-tech-violates-civil-rights-act/}}.


\bibitem[Mollah et~al\mbox{.}(2017)]%
        {mollah}
\bibfield{author}{\bibinfo{person}{Muhammad~Baqer Mollah}, \bibinfo{person}{Md. Abul~Kalam Azad}, {and} \bibinfo{person}{Athanasios Vasilakos}.} \bibinfo{year}{2017}\natexlab{}.
\newblock \showarticletitle{Security and privacy challenges in mobile cloud computing: {Survey} and way ahead}.
\newblock \bibinfo{journal}{\emph{Journal of Network and Computer Applications}}  \bibinfo{volume}{84} (\bibinfo{date}{April} \bibinfo{year}{2017}), \bibinfo{pages}{38--54}.
\newblock
\showISSN{1084-8045}
\urldef\tempurl%
\url{https://doi.org/10.1016/j.jnca.2017.02.001}
\showDOI{\tempurl}


\bibitem[Mostowski and Vullers(2012)]%
        {u-prove}
\bibfield{author}{\bibinfo{person}{Wojciech Mostowski} {and} \bibinfo{person}{Pim Vullers}.} \bibinfo{year}{2012}\natexlab{}.
\newblock \showarticletitle{{Efficient U-Prove Implementation for Anonymous Credentials on Smart Cards}}. In \bibinfo{booktitle}{\emph{Security and Privacy in Communication Networks}} (London) \emph{(\bibinfo{series}{SecureComm 2011})}, \bibfield{editor}{\bibinfo{person}{Muttukrishnan Rajarajan}, \bibinfo{person}{Fred Piper}, \bibinfo{person}{Haining Wang}, {and} \bibinfo{person}{George Kesidis}} (Eds.). \bibinfo{publisher}{Springer Berlin Heidelberg}, \bibinfo{address}{Berlin, Heidelberg}, \bibinfo{pages}{243--260}.
\newblock
\urldef\tempurl%
\url{https://doi.org/10.1007/978-3-642-31909-9_14}
\showDOI{\tempurl}


\bibitem[Nershi(2022)]%
        {cryptocurrency}
\bibfield{author}{\bibinfo{person}{Karen Nershi}.} \bibinfo{year}{2022}\natexlab{}.
\newblock \bibinfo{title}{How Strong Are International Standards in Practice? {E}vidence from Cryptocurrency Transactions}.
\newblock \bibinfo{howpublished}{Stanford Cyber Policy Center Fall Seminar Series}.
\newblock
\newblock
\shownote{\url{https://www.youtube.com/watch?v=XHD-3lUaVUE}}.


\bibitem[Nissenbaum(2011)]%
        {contextualintegrity}
\bibfield{author}{\bibinfo{person}{Helen Nissenbaum}.} \bibinfo{year}{2011}\natexlab{}.
\newblock \showarticletitle{{A Contextual Approach to Privacy Online}}.
\newblock \bibinfo{journal}{\emph{Daedalus}} \bibinfo{volume}{140}, \bibinfo{number}{4} (\bibinfo{date}{Oct.} \bibinfo{year}{2011}), \bibinfo{pages}{32--48}.
\newblock
\urldef\tempurl%
\url{https://doi.org/10.1162/DAED_a_00113}
\showDOI{\tempurl}


\bibitem[of~Motor~Vehicles)(nd)]%
        {ca-dmv}
\bibfield{author}{\bibinfo{person}{California DMV~(Department of Motor~Vehicles)}.} \bibinfo{year}{[n.d.]}\natexlab{}.
\newblock \bibinfo{title}{{CA DMV Wallet \& mDL Pilot}}.
\newblock
\newblock
\newblock
\shownote{\url{https://www.dmv.ca.gov/portal/ca-dmv-wallet/}}.


\bibitem[Page(2009)]%
        {hotel}
\bibfield{author}{\bibinfo{person}{Lewis Page}.} \bibinfo{year}{2009}\natexlab{}.
\newblock \showarticletitle{Facial-recognition tech now used to greet hotel guests}.
\newblock \bibinfo{journal}{\emph{The Register}} (\bibinfo{date}{Feb.} \bibinfo{year}{2009}).
\newblock
\newblock
\shownote{\url{https://www.theregister.com/2009/02/13/face_ware_hotel_guests}}.


\bibitem[Pfitzmann and Hansen(2010)]%
        {minimization}
\bibfield{author}{\bibinfo{person}{Andreas Pfitzmann} {and} \bibinfo{person}{Marit Hansen}.} \bibinfo{year}{2010}\natexlab{}.
\newblock \bibinfo{title}{{A terminology for talking about privacy by data minimization: Anonymity, Unlinkability, Undetectability, Unobservability, Pseudonymity, and Identity Management}}.
\newblock
\newblock
\newblock
\shownote{\url{http://dud.inf.tu-dresden.de/literatur/Anon\_Terminology\_v0.34.pdf}}.


\bibitem[Pichai(2018)]%
        {Google}
\bibfield{author}{\bibinfo{person}{Sundar Pichai}.} \bibinfo{year}{2018}\natexlab{}.
\newblock \bibinfo{title}{AI at Google: our principles}.
\newblock
\newblock
\urldef\tempurl%
\url{https://blog.google/technology/ai/ai-principles/}
\showURL{%
\tempurl}


\bibitem[Powers et~al\mbox{.}(2024)]%
        {powers}
\bibfield{author}{\bibinfo{person}{Carson Powers}, \bibinfo{person}{Nickolas Gravel}, \bibinfo{person}{Christopher Pellegrini}, \bibinfo{person}{Micah Sherr}, \bibinfo{person}{Michelle~L. Mazurek}, {and} \bibinfo{person}{Daniel Votipka}.} \bibinfo{year}{2024}\natexlab{}.
\newblock \showarticletitle{{``I can say I'm John Travolta...but I'm not John Travolta'': Investigating the Impact of Changes to Social Media Verification Policies on User Perceptions of Verified Accounts}}. In \bibinfo{booktitle}{\emph{Twentieth Symposium on Usable Privacy and Security}} \emph{(\bibinfo{series}{SOUPS 2024})}. \bibinfo{publisher}{USENIX Association}, \bibinfo{address}{Philadelphia, PA}, \bibinfo{pages}{353--372}.
\newblock
\urldef\tempurl%
\url{https://www.usenix.org/conference/soups2024/presentation/powers}
\showURL{%
\tempurl}


\bibitem[Rappeport and Hill(2022)]%
        {irs}
\bibfield{author}{\bibinfo{person}{Alan Rappeport} {and} \bibinfo{person}{Kashmir Hill}.} \bibinfo{year}{2022}\natexlab{}.
\newblock \showarticletitle{{I.R.S. to End Use of Facial Recognition for Identity Verification}}.
\newblock \bibinfo{journal}{\emph{The New York Times}} (\bibinfo{date}{7 Feb.} \bibinfo{year}{2022}).
\newblock
\newblock
\shownote{\url{https://www.nytimes.com/2022/02/07/us/politics/irs-idme-facial-recognition.html}}.


\bibitem[Richards and Hartzog(2019)]%
        {pathologies}
\bibfield{author}{\bibinfo{person}{Neil Richards} {and} \bibinfo{person}{Woodrow Hartzog}.} \bibinfo{year}{2019}\natexlab{}.
\newblock \showarticletitle{{The Pathologies of Digital Consent}}.
\newblock \bibinfo{journal}{\emph{Washington University Law Review}}  \bibinfo{volume}{96} (\bibinfo{year}{2019}), \bibinfo{pages}{1461--1503}.
\newblock
\newblock
\shownote{\url{https://papers.ssrn.com/abstract=3370433}}.


\bibitem[Sato(2021)]%
        {sato}
\bibfield{author}{\bibinfo{person}{Mia Sato}.} \bibinfo{year}{2021}\natexlab{}.
\newblock \showarticletitle{The pandemic is testing the limits of face recognition}.
\newblock \bibinfo{journal}{\emph{MIT Technology Review}} (\bibinfo{date}{28 Sept.} \bibinfo{year}{2021}).
\newblock
\newblock
\shownote{\url{https://www.technologyreview.com/2021/09/28/1036279/pandemic-unemployment-government-face-recognition/}}.


\bibitem[Scheuerman et~al\mbox{.}(2021)]%
        {scheuerman}
\bibfield{author}{\bibinfo{person}{Morgan~Klaus Scheuerman}, \bibinfo{person}{Jialun~Aaron Jiang}, \bibinfo{person}{Casey Fiesler}, {and} \bibinfo{person}{Jed~R. Brubaker}.} \bibinfo{year}{2021}\natexlab{}.
\newblock \showarticletitle{A Framework of Severity for Harmful Content Online}.
\newblock \bibinfo{journal}{\emph{Proceedings of the ACM on Human-Computer Interaction}} \bibinfo{volume}{5}, \bibinfo{number}{CSCW2}, Article \bibinfo{articleno}{368} (\bibinfo{date}{Oct.} \bibinfo{year}{2021}), \bibinfo{numpages}{33}~pages.
\newblock
\urldef\tempurl%
\url{https://doi.org/10.1145/3479512}
\showDOI{\tempurl}


\bibitem[Scheuerman et~al\mbox{.}(2019)]%
        {classification}
\bibfield{author}{\bibinfo{person}{Morgan~Klaus Scheuerman}, \bibinfo{person}{Jacob~M. Paul}, {and} \bibinfo{person}{Jed~R. Brubaker}.} \bibinfo{year}{2019}\natexlab{}.
\newblock \showarticletitle{How Computers See Gender: An Evaluation of Gender Classification in Commercial Facial Analysis Services}.
\newblock \bibinfo{journal}{\emph{Proceedings of the ACM on Human-Computer Interaction}} \bibinfo{volume}{3}, \bibinfo{number}{CSCW}, Article \bibinfo{articleno}{144} (\bibinfo{date}{Nov.} \bibinfo{year}{2019}), \bibinfo{numpages}{33}~pages.
\newblock
\urldef\tempurl%
\url{https://doi.org/10.1145/3359246}
\showDOI{\tempurl}


\bibitem[Schouten and Jacobs(2009)]%
        {biometricpassports}
\bibfield{author}{\bibinfo{person}{Ben Schouten} {and} \bibinfo{person}{Bart Jacobs}.} \bibinfo{year}{2009}\natexlab{}.
\newblock \showarticletitle{Biometrics and their use in e-passports}.
\newblock \bibinfo{journal}{\emph{Image and Vision Computing: Special Issue on Multimodal Biometrics}} \bibinfo{volume}{27}, \bibinfo{number}{3} (\bibinfo{date}{Feb.} \bibinfo{year}{2009}), \bibinfo{pages}{305--312}.
\newblock
\showISSN{0262-8856}
\urldef\tempurl%
\url{https://doi.org/10.1016/j.imavis.2008.05.008}
\showDOI{\tempurl}


\bibitem[Shelby et~al\mbox{.}(2023)]%
        {shelby}
\bibfield{author}{\bibinfo{person}{Renee Shelby}, \bibinfo{person}{Shalaleh Rismani}, \bibinfo{person}{Kathryn Henne}, \bibinfo{person}{AJung Moon}, \bibinfo{person}{Negar Rostamzadeh}, \bibinfo{person}{Paul Nicholas}, \bibinfo{person}{N'Mah Yilla-Akbari}, \bibinfo{person}{Jess Gallegos}, \bibinfo{person}{Andrew Smart}, \bibinfo{person}{Emilio Garcia}, {and} \bibinfo{person}{Gurleen Virk}.} \bibinfo{year}{2023}\natexlab{}.
\newblock \showarticletitle{{Identifying Sociotechnical Harms of Algorithmic Systems: Scoping a Taxonomy for Harm Reduction}}. In \bibinfo{booktitle}{\emph{Proceedings of the 2023 AAAI/ACM Conference on AI, Ethics, and Society}} \emph{(\bibinfo{series}{AIES 2023})}. \bibinfo{pages}{723–741}.
\newblock
\urldef\tempurl%
\url{https://doi.org/10.1145/3600211.3604673}
\showDOI{\tempurl}


\bibitem[Shoemaker et~al\mbox{.}(2019)]%
        {shoemaker}
\bibfield{author}{\bibinfo{person}{Emrys Shoemaker}, \bibinfo{person}{Gudrun~Svava Kristinsdottir}, \bibinfo{person}{Tanuj Ahuja}, \bibinfo{person}{Dina Baslan}, \bibinfo{person}{Bryan Pon}, \bibinfo{person}{Paul Currion}, \bibinfo{person}{Pius Gumisizira}, {and} \bibinfo{person}{Nicola Dell}.} \bibinfo{year}{2019}\natexlab{}.
\newblock \showarticletitle{Identity at the margins: examining refugee experiences with digital identity systems in Lebanon, Jordan, and Uganda}. In \bibinfo{booktitle}{\emph{Proceedings of the 2nd ACM SIGCAS Conference on Computing and Sustainable Societies}} \emph{(\bibinfo{series}{COMPASS '19})}. \bibinfo{publisher}{Association for Computing Machinery}, \bibinfo{address}{New York, NY, USA}, \bibinfo{pages}{206–217}.
\newblock
\urldef\tempurl%
\url{https://doi.org/10.1145/3314344.3332486}
\showDOI{\tempurl}


\bibitem[Sporny et~al\mbox{.}(2022)]%
        {W3C}
\bibfield{author}{\bibinfo{person}{Manu Sporny}, \bibinfo{person}{Dave Longley}, \bibinfo{person}{Markus Sabadello}, \bibinfo{person}{Drummond Reed}, \bibinfo{person}{Orie Steele}, {and} \bibinfo{person}{Christopher Allen}.} \bibinfo{year}{2022}\natexlab{}.
\newblock \bibinfo{title}{{Decentralized Identifiers (DIDs) v1.0: Core architecture, data model, and representations}}.
\newblock
\newblock
\newblock
\shownote{\url{https://www.w3.org/TR/did-core/}}.


\bibitem[Stahl and Wright(2018)]%
        {stahl}
\bibfield{author}{\bibinfo{person}{Bernd~Carsten Stahl} {and} \bibinfo{person}{David Wright}.} \bibinfo{year}{2018}\natexlab{}.
\newblock \showarticletitle{{Ethics and Privacy in AI and Big Data: Implementing Responsible Research and Innovation}}.
\newblock \bibinfo{journal}{\emph{IEEE Security \& Privacy}} \bibinfo{volume}{16}, \bibinfo{number}{3} (\bibinfo{year}{2018}), \bibinfo{pages}{26--33}.
\newblock
\urldef\tempurl%
\url{https://doi.org/10.1109/MSP.2018.2701164}
\showDOI{\tempurl}


\bibitem[Stapleberg(2025)]%
        {zero}
\bibfield{author}{\bibinfo{person}{Alan Stapleberg}.} \bibinfo{year}{2025}\natexlab{}.
\newblock \bibinfo{title}{It's now easier to prove age and identity with Google Wallet}.
\newblock
\newblock
\urldef\tempurl%
\url{https://blog.google/products/google-pay/google-wallet-age-identity-verifications/}
\showURL{%
\tempurl}


\bibitem[Staton(2016)]%
        {staton}
\bibfield{author}{\bibinfo{person}{Bethan Staton}.} \bibinfo{year}{2016}\natexlab{}.
\newblock \showarticletitle{Eye spy: biometric aid system trials in {J}ordan}.
\newblock \bibinfo{journal}{\emph{The New Humanitarian}} (\bibinfo{date}{May} \bibinfo{year}{2016}).
\newblock
\newblock
\shownote{\url{https://www.thenewhumanitarian.org/analysis/2016/05/18/eye-spy-biometric-aid-system-trials-jordan}}.


\bibitem[Stokkink et~al\mbox{.}(2021)]%
        {trulySSI}
\bibfield{author}{\bibinfo{person}{Quinten Stokkink}, \bibinfo{person}{Georgy Ishmaev}, \bibinfo{person}{Dick Epema}, {and} \bibinfo{person}{Johan Pouwelse}.} \bibinfo{year}{2021}\natexlab{}.
\newblock \showarticletitle{A Truly Self-Sovereign Identity System}. In \bibinfo{booktitle}{\emph{2021 IEEE 46th Conference on Local Computer Networks (LCN)}}. \bibinfo{pages}{1--8}.
\newblock
\urldef\tempurl%
\url{https://doi.org/10.1109/LCN52139.2021.9525011}
\showDOI{\tempurl}


\bibitem[Temoshok et~al\mbox{.}(2024)]%
        {nistv4}
\bibfield{author}{\bibinfo{person}{David Temoshok}, \bibinfo{person}{Diana Proud-Madruga}, \bibinfo{person}{Yee-Yin Choong}, \bibinfo{person}{Ryan Galluzzo}, \bibinfo{person}{Sarbari Gupta}, \bibinfo{person}{Connie LaSalle}, \bibinfo{person}{Naomi Lefkovitz}, {and} \bibinfo{person}{Andrew Regenscheid}.} \bibinfo{year}{2024}\natexlab{}.
\newblock \showarticletitle{{Digital Identity Guidelines}}.
\newblock \bibinfo{journal}{\emph{NIST Special Publication}}  \bibinfo{volume}{800-63-4 (2nd Public Draft)} (\bibinfo{date}{Aug.} \bibinfo{year}{2024}).
\newblock
\urldef\tempurl%
\url{https://doi.org/10.6028/NIST.SP.800-63-4.2pd}
\showDOI{\tempurl}


\bibitem[Thomas(2006)]%
        {thomas}
\bibfield{author}{\bibinfo{person}{David~R. Thomas}.} \bibinfo{year}{2006}\natexlab{}.
\newblock \showarticletitle{A General Inductive Approach for Analyzing Qualitative Evaluation Data}.
\newblock \bibinfo{journal}{\emph{American Journal of Evaluation}} \bibinfo{volume}{27}, \bibinfo{number}{2} (\bibinfo{year}{2006}), \bibinfo{pages}{237--246}.
\newblock
\urldef\tempurl%
\url{https://doi.org/10.1177/1098214005283748}
\showDOI{\tempurl}


\bibitem[Totilo(2021)]%
        {totilo}
\bibfield{author}{\bibinfo{person}{Stephen Totilo}.} \bibinfo{year}{2021}\natexlab{}.
\newblock \showarticletitle{Youth gaming platform {R}oblox rolls out new age verification system}.
\newblock \bibinfo{journal}{\emph{Axios}} (\bibinfo{date}{21 Sept.} \bibinfo{year}{2021}).
\newblock
\newblock
\shownote{\url{https://www.axios.com/2021/09/21/roblox-age-youth-identity-verification-gaming}}.


\bibitem[Valle(2024)]%
        {delvalle}
\bibfield{author}{\bibinfo{person}{Gaby~Del Valle}.} \bibinfo{year}{2024}\natexlab{}.
\newblock \showarticletitle{Don’t ever hand your phone to the cops}.
\newblock \bibinfo{journal}{\emph{The Verge}} (\bibinfo{date}{Sept.} \bibinfo{year}{2024}).
\newblock
\newblock
\shownote{\url{https://www.theverge.com/2024/9/24/24252235/police-unlock-phone-password-face-id-apple-wallet-id}}.


\bibitem[Vullers and Alp{\'a}r(2013)]%
        {selective}
\bibfield{author}{\bibinfo{person}{Pim Vullers} {and} \bibinfo{person}{Gergely Alp{\'a}r}.} \bibinfo{year}{2013}\natexlab{}.
\newblock \showarticletitle{{Efficient Selective Disclosure on Smart Cards Using Idemix}}. In \bibinfo{booktitle}{\emph{Policies and Research in Identity Management}}, \bibfield{editor}{\bibinfo{person}{Simone Fischer-H{\"u}bner}, \bibinfo{person}{Elisabeth de~Leeuw}, {and} \bibinfo{person}{Chris Mitchell}} (Eds.). \bibinfo{publisher}{Springer Berlin Heidelberg}, \bibinfo{address}{Berlin, Heidelberg}, \bibinfo{pages}{53--67}.
\newblock
\urldef\tempurl%
\url{https://doi.org/10.1007/978-3-642-37282-7_5}
\showDOI{\tempurl}


\bibitem[Wang and De~Filippi(2020)]%
        {economicinclusion}
\bibfield{author}{\bibinfo{person}{Fennie Wang} {and} \bibinfo{person}{Primavera De~Filippi}.} \bibinfo{year}{2020}\natexlab{}.
\newblock \showarticletitle{Self-sovereign identity in a globalized world: Credentials-based identity systems as a driver for economic inclusion}.
\newblock \bibinfo{journal}{\emph{Frontiers in Blockchain}}  \bibinfo{volume}{2} (\bibinfo{date}{Jan.} \bibinfo{year}{2020}).
\newblock
\urldef\tempurl%
\url{https://doi.org/10.3389/fbloc.2019.00028}
\showDOI{\tempurl}


\bibitem[Warford et~al\mbox{.}(2022)]%
        {warford}
\bibfield{author}{\bibinfo{person}{Noel Warford}, \bibinfo{person}{Tara Matthews}, \bibinfo{person}{Kaitlyn Yang}, \bibinfo{person}{Omer Akgul}, \bibinfo{person}{Sunny Consolvo}, \bibinfo{person}{Patrick~Gage Kelley}, \bibinfo{person}{Nathan Malkin}, \bibinfo{person}{Michelle~L Mazurek}, \bibinfo{person}{Manya Sleeper}, {and} \bibinfo{person}{Kurt Thomas}.} \bibinfo{year}{2022}\natexlab{}.
\newblock \showarticletitle{{SoK: A Framework for Unifying At-Risk User Research}}. In \bibinfo{booktitle}{\emph{43rd IEEE Symposium on Security and Privacy}} \emph{(\bibinfo{series}{Oakland '22})}. IEEE, \bibinfo{pages}{2344--2360}.
\newblock
\urldef\tempurl%
\url{https://doi.org/10.1109/SP46214.2022.9833643}
\showDOI{\tempurl}


\bibitem[Watkins(2021)]%
        {watkins}
\bibfield{author}{\bibinfo{person}{Elizabeth~Anne Watkins}.} \bibinfo{year}{2021}\natexlab{}.
\newblock \bibinfo{title}{{``In the Pre-Evening Light it Sees So Little'': A Human-Centered, Qualitative Investigation into Computer Vision as Account Verification in Gig Work}}.
\newblock \bibinfo{howpublished}{Princeton CITP Seminar}.
\newblock
\newblock
\shownote{\url{https://www.cs.princeton.edu/events/26112}}.


\bibitem[Watkins(2023)]%
        {watkinsCSCW}
\bibfield{author}{\bibinfo{person}{Elizabeth~Anne Watkins}.} \bibinfo{year}{2023}\natexlab{}.
\newblock \showarticletitle{Face Work: A Human-Centered Investigation into Facial Verification in Gig Work}.
\newblock \bibinfo{journal}{\emph{Proceedings of the ACM on Human-Computer Interaction}} \bibinfo{volume}{7}, \bibinfo{number}{CSCW1}, Article \bibinfo{articleno}{52} (\bibinfo{date}{April} \bibinfo{year}{2023}), \bibinfo{numpages}{24}~pages.
\newblock
\urldef\tempurl%
\url{https://doi.org/10.1145/3579485}
\showDOI{\tempurl}


\bibitem[Weidinger et~al\mbox{.}(2023)]%
        {weidinger}
\bibfield{author}{\bibinfo{person}{Laura Weidinger}, \bibinfo{person}{Maribeth Rauh}, \bibinfo{person}{Nahema Marchal}, \bibinfo{person}{Arianna Manzini}, \bibinfo{person}{Lisa~Anne Hendricks}, \bibinfo{person}{Juan Mateos-Garcia}, \bibinfo{person}{Stevie Bergman}, \bibinfo{person}{Jackie Kay}, \bibinfo{person}{Conor Griffin}, \bibinfo{person}{Ben Bariach}, \bibinfo{person}{Iason Gabriel}, \bibinfo{person}{Verena Rieser}, {and} \bibinfo{person}{William Isaac}.} \bibinfo{year}{2023}\natexlab{}.
\newblock \bibinfo{title}{{Sociotechnical Safety Evaluation of Generative AI Systems}}.
\newblock
\newblock
\showeprint[arxiv]{2310.11986}~[cs.AI]


\bibitem[Whitley et~al\mbox{.}(2014)]%
        {whitley2014}
\bibfield{author}{\bibinfo{person}{Edgar~A. Whitley}, \bibinfo{person}{Uri Gal}, {and} \bibinfo{person}{Annemette Kjaergaard}.} \bibinfo{year}{2014}\natexlab{}.
\newblock \showarticletitle{Who do you think you are? A review of the complex interplay between information systems, identification and identity}.
\newblock \bibinfo{journal}{\emph{European Journal of Information Systems}}  \bibinfo{volume}{23} (\bibinfo{year}{2014}), \bibinfo{pages}{17--35}.
\newblock
\urldef\tempurl%
\url{https://doi.org/10.1057/ejis.2013.34}
\showDOI{\tempurl}


\bibitem[Whittaker(2021)]%
        {palmprint}
\bibfield{author}{\bibinfo{person}{Zack Whittaker}.} \bibinfo{year}{2021}\natexlab{}.
\newblock \showarticletitle{Amazon will pay you \$10 in credit for your palm print biometrics}.
\newblock \bibinfo{journal}{\emph{TechCrunch}} (\bibinfo{date}{Aug.} \bibinfo{year}{2021}).
\newblock
\newblock
\shownote{\url{https://techcrunch.com/2021/08/02/amazon-credit-palm-biometrics}}.


\bibitem[Woodruff(2019)]%
        {tenthings}
\bibfield{author}{\bibinfo{person}{Allison Woodruff}.} \bibinfo{year}{2019}\natexlab{}.
\newblock \showarticletitle{10 things you should know about algorithmic fairness}.
\newblock \bibinfo{journal}{\emph{Interactions}} \bibinfo{volume}{26}, \bibinfo{number}{4} (\bibinfo{date}{June} \bibinfo{year}{2019}), \bibinfo{pages}{47–51}.
\newblock
\showISSN{1072-5520}
\urldef\tempurl%
\url{https://doi.org/10.1145/3328489}
\showDOI{\tempurl}


\bibitem[Zucker-Scharff(2022)]%
        {mta}
\bibfield{author}{\bibinfo{person}{Aram Zucker-Scharff}.} \bibinfo{year}{2022}\natexlab{}.
\newblock \showarticletitle{The {MTA}'s switch to {OMNY} machines is a privacy nightmare}.
\newblock \bibinfo{journal}{\emph{Fast Company}} (\bibinfo{date}{Sept.} \bibinfo{year}{2022}).
\newblock
\newblock
\shownote{\url{https://www.fastcompany.com/90788367/the-mtas-switch-to-omny-machines-is-a-privacy-nightmare}}.


\end{thebibliography}


\end{document}